\documentclass[fleqn,usenatbib]{mnras}

\usepackage[T1]{fontenc}
\usepackage{ae,aecompl}

\usepackage{graphicx}	
\usepackage{amsmath}	
\usepackage{amssymb}	

\newcommand{\D}{\mathrm{d}}
\newcommand{\Ms}{{\ensuremath{\mathrm{M}_{\sun}}}}

\newcommand{\Tc}{{\ensuremath{{T}_\mathrm{crit}}}}
\newcommand{\K}{{\ensuremath{\mathrm{K}}}}
\newcommand{\Myr}{{\ensuremath{\mathrm{Myr}}}}
\newcommand{\Mpc}{{\ensuremath{\mathrm{Mpc}}}}
\newcommand{\fII}{{\ensuremath{f_\mathrm{II}}}}

\newcommand{\Lya}{Ly-${\alpha}$}

\title[The cosmological transition to metal-enriched star-formation]{Effect of the cosmological transition to metal-enriched star-formation on the hydrogen 21-cm signal}

\author[M. Magg et al.]{Mattis Magg$^{1,2}$\thanks{E-mail: mattis.magg@protonmail.com}, Itamar Reis$^{3}$, Anastasia Fialkov$^{4,5}$, Rennan Barkana$^{3}$, 
\newauthor Ralf S. Klessen$^{1,6}$, Simon C. O. Glover$^1$, Li-Hsin Chen$^{1,2}$, Tilman Hartwig$^{7,8,9}$
\newauthor Anna T.P. Schauer$^{10}$\thanks{NHFP Hubble Fellow}\\
$^{1}$Universit\"at Heidelberg, Zentrum f\"ur Astronomie, Institut f\"ur Theoretische Astrophysik, D-69120 Heidelberg, Germany\\
$^{2}$International Max Planck Research School for Astronomy and Cosmic Physics at the University of Heidelberg (IMPRS-HD)\\
$^{3}$School of Physics and Astronomy, Tel-Aviv University, Tel-Aviv, 69978, Israel\\
$^{4}$Institute of Astronomy, University of Cambridge, Madingley Road, Cambridge, CB3 0HA, UK\\
$^{5}$Kavli Institute for Cosmology, Madingley Road, Cambridge, CB3 0HA, UK\\
$^{6}$Universit\"{a}t Heidelberg, Interdiszipli\"{a}res Zentrum f\"{u}r Wissenschaftliches Rechnen, D-69120 Heidelberg, Germany\\
$^{7}$Department of Physics, School of Science, The University of Tokyo, Bunkyo, Tokyo 113-0033, Japan\\
$^{8}$Institute for Physics of Intelligence, School of Science, The University of Tokyo, Bunkyo, Tokyo 113-0033, Japan\\
$^{9}$Kavli Institute for the Physics and Mathematics of the Universe (WPI), The University of Tokyo Institutes for Advanced Study,\\
The University of Tokyo, Kashiwa, Chiba 277-8583, Japan\\
$^{10}$Department of Astronomy, University of Texas at Austin, Austin, TX 78712, USA\\}

\date{Accepted XXX. Received YYY; in original form ZZZ}

\pubyear{2021}

\begin{document}
\label{firstpage}
\pagerange{\pageref{firstpage}--\pageref{lastpage}}
\maketitle

\begin{abstract}
Mapping Cosmic Dawn with 21-cm tomography offers an exciting new window into the era of primordial star formation. However, self-consistent implementation of both the process of star formation and the related 21-cm signal is challenging, due to the multi-scale nature of the problem. In this study, we develop a flexible semi-analytical model to follow the formation of the first stars and the process of gradual transition from primordial to metal-enriched star formation. For this transition we use different in scenarios with varying time-delays (or recovery times) between the first supernovae and the formation of the second generation of stars. We use recovery times between 10 and 100\,Myr and find that these delays have a strong impact on the redshift at which the transition to metal-enriched star formation occurs. We then explore the effect of this transition on the 21-cm signal and find that the recovery time has a distinctive imprint in the signal. Together with an improved understanding of how this time-delay relates to the properties of Population~III stars, future 21-cm observations can give independent constraints on the earliest epoch of star formation.
\end{abstract} 



\section{INTRODUCTION}
The first stars (Population~III or Pop~III stars) are expected to start forming around 30 million years after the Big Bang \citep{Naoz2006, Fialkov12}. As by definition they form from pristine material, unaffected by previous generations of stars, their birth clouds are metal-free. This absence of metals reduces the possibilities for gas cooling and results in much higher gas temperatures in the star-forming regions of the early Universe, which leads to the formation of comparatively massive stars \citep{Bromm2002, Yoshida2003, GloverReview, GreifReview, Hosokawa16}. The first stars enrich their surroundings with heavy elements and enhance the cooling capability of the gas and therefore cause the transition to low mass (Population~II or Pop~II stars) star formation \citep[][]{Bromm2001, Schneider2003, Ritter12, Jeon15, Chiaki19}. The metallicity at which this transition occurs is still subject to debate and potentially is sensitive to the elemental composition of the ejected gas and the properties of the dust grains within it \citep{Omukai05, Schneider12, Chiaki15}.

Properties of the first stars are still poorly constrained. Early simulations suggested typical masses between one hundred and several thousands of solar masses \citep[e.g.][]{Abel02, Omukai03, Bromm2002}, whereas later simulations suggest much lower stellar masses \citep{clark11, Greif11b, Stacy16}. However, the absence of detections of metal-free stars until today shows that their initial mass function (IMF) must be different from the one found in the present day Universe \citep{Salvadori07, Magg19, Rossi2021}, and likely consisted of more massive stars.

As of now, the epoch in which these stars form is still relatively inaccessible to astronomy. Most observational studies investigating the first stars use an indirect method, namely stellar archaeology \citep{Beers2005, Frebel15}. In this approach the elemental abundance patterns observed in metal-poor stars \citep[e.g.][]{keller14, caffau12, Nordlander2019} are compared to models of the first supernovae \citep[SNe; e.g.][]{Nomoto13, HegerWoosley2010}. This comparison can then shed light on the typical properties of the first SNe and therefore their progenitor stars \citep{Fraser17, Ishigaki18, Ezzeddine19}. 

Observations of the cosmic 21-cm signal offer a new alternative way to probe the onset of star formation and constrain the properties of Pop~III and Pop~II stars. The cosmic 21-cm signal is produced by neutral inter-galactic hydrogen in the high redshift Universe \citep[see][for recent reviews of the topic]{barkana18book, Mesingerbook}. It provides a window to the evolution of the Universe between the Dark Ages ($z\sim100$) through Cosmic Dawn ($z\sim15-25$) until the Epoch of Reionization (EoR, $z\sim6-10$). The 21-cm signal is predominantly determined by the occupancy of the hydrogen hyperfine levels (characterized by the spin temperature, which depends on the luminosity of high-redshift sources in ultraviolet, X-ray and radio bands), the ionization state of the gas as well as its density and velocity \citep[e.g.][]{madau97}. Owing to its dependence on the properties of sources, the 21-cm signal can be used to characterize high redshift stars, black holes and galaxies at high redshift.

Observational effort to detect both the sky-averaged (global) 21-cm signal and its fluctuations across the sky is ongoing. Experiments aiming to measure the global signal include EDGES \citep{bowman13}, LEDA \citep{Price:2018}, SARAS \citep{Singh:2018}, PRIZM \citep{philip19}, MIST\footnote{\url{http://www.physics.mcgill.ca/mist/}} and REACH\footnote{\url{https://www.kicc.cam.ac.uk/projects/reach}}; while interferometers including the LOFAR \citep{Gehlot:2019}, HERA \citep{deboer17}, LWA \citep{Eastwood:2019}, NenuFAR \citep{zarka12}, LEDA \citep{Garsden:2021} and MWA \citep{MWA} as well as the future SKA \citep{koopmans15} aim to measure the fluctuations of the 21-cm signal from the EoR and Cosmic Dawn. The first tentative detection of the global signal was reported by the EDGES collaboration \citep{EDGES18}. Although the true nature of this signal is still debated \citep[e.g.\ see][]{Hills:2018, Sims:2020}, if it truly is of cosmological origin it represents a direct evidence of star formation at $z\sim 17$ when the Universe was $\sim 230$ million years old.

\begin{figure}
 \includegraphics[width=\linewidth]{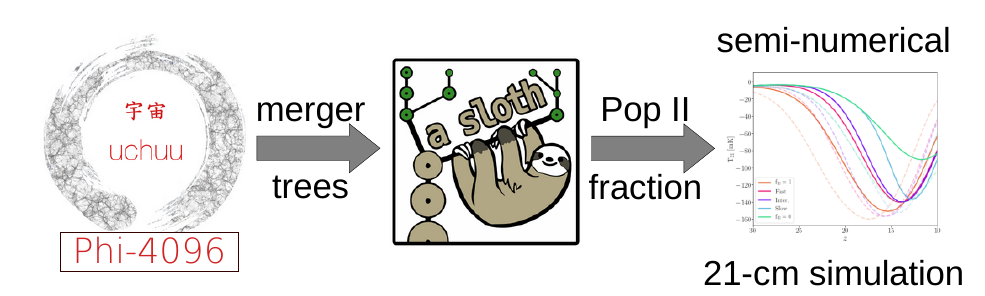}
 \caption{\label{fig:workflow}Illustration of our workflow. The Uchuu simulations and the generation of merger trees are part of \citet{Ishiyama2021}.}
\end{figure}

Here, we aim at investigating the effect of the transition between the first- and second generation stars (Pop~III and Pop~II stars) on the global 21-cm signal and the power spectrum of 21-cm fluctuations. This effect is inherently difficult to model, since the minihaloes in which the first stars form have sizes of the order of one comoving kpc, yet due to the large mean free paths of ultraviolet and X-ray radiation, the 21cm signal is shaped on scales of hundreds of comoving Mpc. It is not yet feasible to simulate the effect of first stars on the large-scale 21-cm signal in a self-consistent way. Therefore, we use the following approach: We model the formation of Pop~III and Pop~II stars with the semi-analytical model \textsc{a-sloth} as described in Section \ref{sect:asloth}. In Section \ref{sect:fitting} we use those results to model the transition from metal-free to metal-enriched star formation under varying sets of assumptions for the recovery time, i.e., the time-delay between the first SNe and the formation of the first Pop~II stars, the local large-scale density field and the minimum mass of star forming haloes. This prescription is then used in large-scale semi-numerical cosmological simulations of the 21-cm signal (Section \ref{Sec:21cmSims}) with the results discussed in Section \ref{sec:res}. Since our workflow relies on three different simulation methods, we include a schematic representation in Fig. \ref{fig:workflow}. We discuss our results in the context of other existing works in the field in Section \ref{sec:disc}. Finally, we conclude in Section \ref{sec:conc}.

\section{Semi-analytical prescription for PopIII-PopII transition}
 \label{sect:asloth}
 \subsection{$N$-body simulations}
 \label{sect:nbody}
We base our semi-analytical model on merger trees generated from the \textit{Uchuu} cosmological dark-matter-only $N$-body simulations \citep{Ishiyama2021}, specifically the simulations labelled Phi-4096. We briefly summarize how the $N$-body simulations are set up and how the merger trees were generated. More details can be found in \citep{Ishiyama2021}. The Phi-4096 simulation models the formation of structure in a box with a comoving edge-length of 16\,Mpc\,$h^{-1}$ and a resolution of $4096^3$ particles, corresponding to a particle mass of $5.13\times10^3\,\Ms\,h^{-1}$. Having this high resolution is important as it allows us to follow even the smallest haloes in which stars may form. Initial conditions were generated with \textsc{music} \citep{music} and use the cosmological parameters from \citet{Planck2013}: $\Omega_m=0.31, \Omega_b=0.048, \Omega_\Lambda=0.69, h=0.68, n_s=0.96$ and $\sigma_8=0.83$. The simulations are initialized at $z_\mathrm{ini}=127$. 

Snap-shots are spaced regularly in intervals of $\Delta \log_{10}\frac{1}{z+1} = 0.01$, where the first snapshot is at $z\approx31$ and the last snapshot for which the merger-trees are available is at $z\approx10$. This leads to the time between the snapshots being 3.5\,Myr at the highest and 16\,Myr at the lowest redshifts. The halo properties were computed with \textsc{rockstar} \citep{rockstar} and the merger trees were generated with \textsc{consistent-trees} \citep{ConsistentTrees}. At a minimum of 40 particles per halo, the merger trees contain haloes with a minimum mass of $2\,\times10^5\,\Ms\,h^{-1}$ and consist of a total of 1.3 billion tree-nodes.

\subsection{Basic semi-analytical model}
In order to quantify the transition from metal-free to metal-enriched star formation we use \textsc{a-sloth}\footnote{Ancient Stars and Local Observables by Tracing Haloes} (Magg et al. in prep). This semi-analytical model simulates early star formation based on merger trees which in our case are taken from the $N$-body simulations described in Section \ref{sect:nbody}. The model is based on the premise that only haloes with masses above a threshold value $M_\mathrm{crit}$ (minimum cooling mass) will form the first (i.e., Pop~III) stars and will subsequently transition to forming Pop~II stars after the first SNe that enrich the gas with metals. We here aim at parametrizing the transition from metal-free to metal-enriched star formation, in different cosmic overdensities as a function of the critical mass.

The criterion to decide when haloes form stars for the first time is parametrized by $\Tc$, which is the critical virial temperature of the halo, related to $ M_\mathrm{crit}$ via
\begin{equation}
 M_\mathrm{crit} = 1.0\times 10^6\,\Ms \left(\frac{\Tc}{10^3\,\K} \right)^{3/2} 
 \left( \frac{1+z}{10}\right)^{-3/2}.
\end{equation}
A typical value for enabling collapse by molecular hydrogen cooling is $\Tc = 2200\,\K$ \citep{Hummel12}.

In contrast to previous models, when using \textsc{a-sloth} we do not explicitly account for the effect of Lyman-Werner (LW) radiation \citep{Machacek2001, Oshea08, Schauer21} or supersonic baryon streaming \citep[a residual velocity difference between dark matter and baryonic matter resulting from recombination,][]{Tseliakhovich10} which are expected to affect $\Tc$ in a non-uniform manner \citep[e.g.][]{ Fialkov12, Schauer19a}. These two effects are taken into account at a later stage (in the cosmological 21-cm simulations described in Section \ref{Sec:21cmSims}. In the semi-analytical model described in this Section we consider \Tc\ to be a free parameter.

We test 29 different values for \Tc, spaced regularly in log-space between $\Tc=1500\,\K$ and $\Tc=50000\,\K$, which covers the range relevant for the early star formation (see Section~\ref{Sec:21cmSims}). The lower limit here roughly corresponds to the smallest haloes that can be resolved with the $N$-body simulations at $z=30$. The upper limit is well above the atomic cooling limit ($\Tc=10000\,\mathrm{K}$), where haloes start to cool efficiently by atomic hydrogen emission. In this regime molecular hydrogen is no longer required to enable cooling, and, therefore, haloes can only be prevented from collapse in relatively extreme conditions \citep{Visbal16, Agarwal2016}. Specifically, \citet{Visbal16} find that haloes as massive as 10 times the atomic cooling limit can be prevented from collapse in the presence of a strong ionizing radiation field, but haloes at even higher masses can not. For this reason we do not consider $\Tc$ values above the $50000\,\mathrm{K}$ threshold. 
 
When a halo first exceeds the mass threshold, it forms Pop~III stars in a single instantaneous burst. In our model, only one generation of metal-free stars can form in each halo, which is what is generally seen in simulations of the first SNe \citep{Jeon14, bsmith15, Chiaki16}. To determine the Pop~III stellar mass we assume that stars form with a fixed star formation efficiency, i.e., when a halo forms Pop~III stars, the stars are sampled from an IMF until they reach a total mass of
 \begin{equation}
 M_{*,\mathrm{III}} = 0.01 \frac{\Omega_\mathrm{b}}{\Omega_\mathrm{m}} M_\mathrm{vir},
 \label{eq:MpopIII}
 \end{equation}
 where the $0.01$ represents our adopted Pop~III star formation efficiency. The stars are sampled from an IMF with the shape
 \begin{equation}
   \frac{\D N}{\D \log(M)} \propto M^{0.5}
 \end{equation}
 within the limits $M \in [2.0\,\Ms, 180\,\Ms]$. Star formation efficiency and IMF are taken from \citet{Tarumi20} and have been calibrated with the metal-poor end of the metallicity distribution function of the Milky Way. In this calibration parameters of Pop~III and Pop~II formation are chosen to match such that the modelled metallicity distribution of metal-poor stars matches the observed distribution. The lifetimes of the stars are taken from \citet{Schaerer2002} and stars within the mass ranges of 8--40\,\Ms\ and 140--260\,\Ms\ are assumed to explode as SN at the end of their lifetime \citep{HegerWoosley2002}. In our main model considered here, the treatment of feedback is highly simplified, and therefore the IMF and star formation efficiency have a very small effect. To be precise, they slightly affect the delay between star formation and the first SNe, because they change the distribution of stellar life-times before the SNe explode and how completely this distribution is sampled. Other effects that can depend on the IMF and star formation efficiency, such as a changed radiation output or a differences in outflows caused by SNe, are not taken into account in this simplified model. In order to avoid artifacts due to the time-discretization of the merger trees \citep[see e.g.][]{Magg16} we assign the time of star formation of a halo randomly between the current time-step and the next.
 
 After the first stars explode, their host halo is enriched with metals and, thus, in principle is able to form metal-enriched Pop~II stars. However, the SNe also eject a very significant amount of energy into the system, heating up the gas and potentially even destroying the halo. The time a halo needs to recover from such a SN is poorly understood and potentially depends on the halo mass and the type and the number of SNe exploding in the halo \citep{Jeon14, Chiaki18}. Such dependencies are not clear yet and especially the statistical scatter between equal mass haloes is poorly understood. Therefore, we assume that the recovery time, $t_\mathrm{recov}$, i.e., the time between the SNe of the first stars and the time of formation of the first Pop~II star in a halo, is a free parameter and is equal for all haloes. We explore the effect of the recovery time by adopting three different values: $t_\mathrm{recov}=10\,\Myr$, $t_\mathrm{recov}=30\,\Myr$ and $t_\mathrm{recov}=100\,\Myr$ to which we refer as \textit{fast}, \textit{intermediate} and \textit{slow} transitions. These values encapsulate the range of values measured in hydrodynamical simulations \citep{Greif10, Jeon14, bsmith15, Chiaki18}. A slower transition does not lead to more Pop~III stars forming, it only means that the formation of the first metal-enriched generation of stars is delayed by a longer period of time. 
 
 In this picture, the slow transition with the longest recovery time is associated with the dominance of small star-forming haloes populated by massive stars, which explode as very energetic SNe and evacuate most of the gas out of the halo. In such a scenario, a long time is required for the haloes to re-collapse and start forming stars again, this time out of the metal-enriched gas \citep{Whalen08b, Jeon14, Chiaki18}. Several SNe going off in the same halo can have a similar effect \citep{Ritter15}. 
 
 The intermediate and the fast transitions correspond to the cases in which the stellar feedback is not strong enough to fully destroy the host haloes, and therefore some baryonic material remains dense and bound to the halo. \citet{Chiaki18} refer to an extreme case of such scenarios as the `inefficient internal enrichment' channel. Whether and how much gas remains in the haloes depends not only on the properties of Pop~III stars, but also on the baryonic substructure of the star-forming haloes. Therefore, while small recovery times could qualitatively be associated with weak feedback, less massive Pop~III stars and a low star formation efficiency, the exact correspondence between $t_\mathrm{recov}$ and the properties of Pop~III stars in this regime is unclear.
 
We model the transition to metal-enriched star formation in two different ways: with our main simplified model and with an additional complete model. In our main model, we neglect the back-reaction that photoionization feedback and external enrichment would have on the Pop~III fractions\footnote{While the impact of reionization at the time of the Pop~III-Pop~II transition is expected to be small as most of the Universe is neutral at high redshifts, we do account for the ionizing feedback in the large-scale 21-cm simulations described in Section \ref{Sec:21cmSims}.}. These simplifications are well justified for two reasons: Firstly, the external metal enrichment is expected to have only a small effect at the high redshifts of interest \citep{Visbal18, Visbal20}. Secondly, the Universe is expected to be largely neutral at these times. We verify that the back-reaction indeed has a negligible effect on the Pop~III-Pop~II transition by exploring an additional `full' model in which these types of feedback are taken into account (see Appendix \ref{apx:feed}). We find that these types of feedback do not have a significant effect on the resulting Pop~II fractions, i.e., that their impact on the Pop~III-Pop~II transition is small compared to the change introduced by varying the recovery time. Finally, we note that these types of feedback may still be important for sub-haloes of larger objects. However, at the moment sub-haloes are not treated in the large-scale 21-cm simulations described in Section \ref{Sec:21cmSims}, which rely on a star formation prescription based on analytical halo mass function

 \subsection{Transition to metal-enriched star formation}
 \label{sect:fitting}
In our semi-analytical model haloes are labelled as Pop~II forming one recovery time after they experienced their first Pop~III SN. To quantify the transition to Pop~II star formation we compare the sum of the (virial) masses of all Pop~II forming haloes with the total mass of all haloes above the star formation threshold. We define the Pop~II fraction, \fII, as the ratio of these two masses, i.e., as the mass fraction of haloes above the critical mass that form Pop~II stars. 

This definition is chosen on purpose over, e.g., considering the stellar masses of Pop~II or Pop~III stars, because, to first order, it is independent of the star formation efficiencies. This feature makes it easy to integrate \fII\ into the large-scale simulations of the 21-cm signal (Section \ref{Sec:21cmSims}). The only way the star formation efficiencies affect the Pop~II fraction in this definition is via the timing of Pop~III SNe (and via the minor effects of external enrichment and radiative feedback, which are not considered in our main model but only in the model in Appendix \ref{apx:feed}). There also is an expected dependence of the recovery time on the Pop~III star formation efficiency, as more stars in a halo should lead to a larger number of SNe and, thus, more efficiently disrupt the halo. However, \citet{Chiaki18} have shown that even two very similar haloes with the same stellar mass content can have vastly different recovery times. Therefore, we assume that the recovery time is independent of the star formation efficiency and treat it as a free parameter.

Driven by the requirements of the large-scale 21-cm simulation (Section \ref{Sec:21cmSims}), we calculate \fII\ in cubic sub-volumes of 3\,Mpc side length, which we will refer to as \textit{pixels} for the remainder of the text. These pixels correspond to the resolution elements of the large-scale 21-cm simulation. As the box has a side length $16\,\mathrm{Mpc}\,h^{-1}=23.52\,\Mpc$, $7^3=343$ pixels can be fit into the box. However, there is an arbitrary choice of placement of the origin of this grid. Therefore, we start the grid at 0,1 and 2\,\Mpc\ from the (0,0,0) corner of the $N$-body simulations along each axis, which leads to 27 grids with a total of 9261 pixels. While this results in a larger range of overdensities and reduces the systematic effect of the arbitrary grid-placement, it introduces a degree of correlation between the pixels. This is an acceptable trade-off, as we do not rely on the pixels being statistically independent in our analysis.

\begin{figure}
 \includegraphics[width=\linewidth]{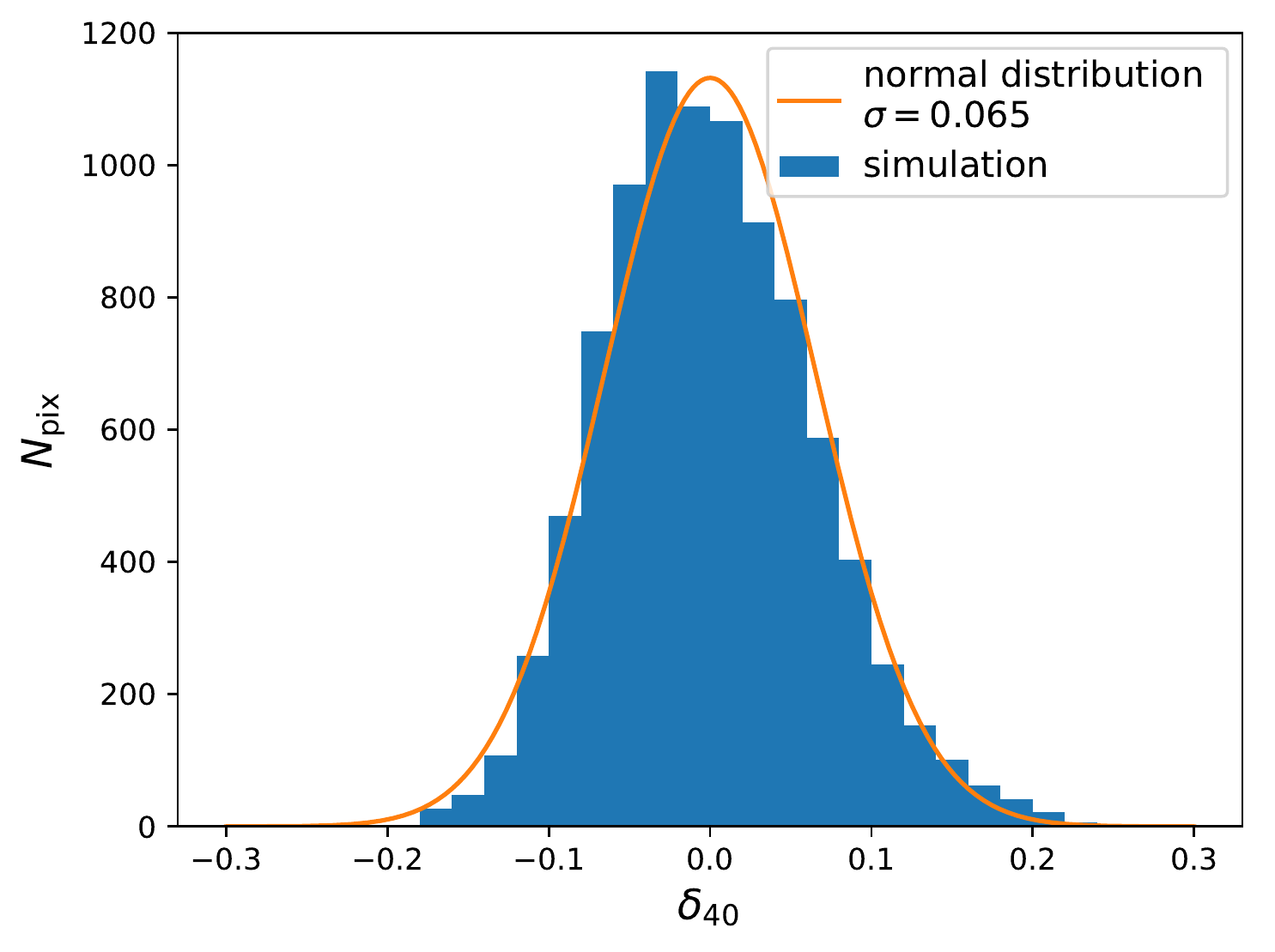}
 \caption{\label{fig:delta_hist}Distribution of rescaled overdensities for all 9261 pixels. The best-fitting Gaussian distribution is shown with an orange line. Despite the slight asymmetry, the distribution is statistically consistent with the Gaussian model.}
\end{figure}

We find that, apart from the obvious dependence on redshift that stems from the growth of structure, the Pop~II fraction depends on the critical virial temperature $\Tc$ as well as on the local overdensity $\delta$ defined as 
\begin{equation}
 \delta = \frac{\rho - \langle\rho\rangle}{\langle\rho\rangle},
 \end{equation}
where $\rho$ is the matter density (here averaged over the size of the pixel) and $\langle\rho\rangle$ is the mean cosmic density calculated across the whole box.
In the hierarchical picture of structure formation, such as we adopt here, star formation in overdense regions will happen earlier than in underdense regions. Consequently, we expect the transition between Pop~III and Pop~II star formation to happen in overdense regions first. Since on the scale of individual pixels the overdensity evolves linearly for the redshift range considered here, we use the density field at $z=40$ as a parameter for fitting (see Eq. \ref{eq:fitting_function}). We compute the overdensitiy within each pixel from the initial conditions and rescale them to $z=40$ using the linear relation
\begin{equation}
 \delta_{40} = \delta \left(\frac{z_\mathrm{ini}+1}{41} \right).
\end{equation}
In Fig. \ref{fig:delta_hist} we present the distribution of overdensities of all the 9261 pixels. We also show the best-fitting Gaussian distribution
\begin{equation}
 f_\mathrm{G}(\delta_{40}, A, \sigma) = N\,\exp\left(-\frac{\delta_{40}^2}{2\sigma^2}\right)
\end{equation}
where $N$ is a normalization parameter and $\sigma=0.065$ is the best-fitting standard deviation. We verified with a Kolmogorov-Smirnov test that despite the visible asymmetry in the distribution of overdensities, the fitted distribution is consistent with the data from the $N$-body simulation. 
 
Next, we fit the following 7-parameter function to the simulated Pop~II fraction:
 \begin{equation}
 \fII = 
 \begin{cases}
 F(z, \delta_{40}, T_3) & \mathrm{if}\ 0 \le F(z, \delta_{40}, T_3) \le 1,\\
 1 & \mathrm{if}\ F(z, \delta_{40}, T_3) > 1,\\
 0 & \mathrm{if}\ F(z, \delta_{40}, T_3) < 0.
 \end{cases}
 \label{eq:fitting_function}
 \end{equation}
 We explicitly impose the physical limits that the Pop~II fraction should be between 0 and 1. $F(z, \delta_{40}, T_3)$ is a function depending on the redshift $z$, the local overdensity averaged over the size of the pixel and normalized to redshift $z=40$, $\delta_{40}$, and the critical virial temperature for star formation expressed as $T_3 = \log_{10} ( \Tc)-3$. We chose the functional form
 \begin{equation}
  F(z, \delta_{40}, T_3) = F_0 +A \arctan\left(\frac{z_0(\delta_{40}, T_3)-z}{\Delta z}\right).
  \label{Eq:func}
 \end{equation}
 Here 
 \begin{equation}
 z_0(\delta_{40}, T_3) = a_2-a_1 T_3-(a_3 T_3+a_4)\delta_{40}
 \label{eq:z_0}
 \end{equation}
 is the redshift of the Pop~III-Pop~II transition and $\Delta z$ is the duration of the transition in units of redshift. Defined in such a way, the transition redshift marks the inflection point in the evolution of the Pop~II fraction (rather than e.g. the half-way point). For our data we find that, typically, at $z_0$ the Pop~II fraction is close to 40 per cent. For convenience, we additionally define a characteristic transition redshift as 
 \begin{equation}
   z_\mathrm{t} = z_0 (\Tc = 2200\,\mathrm{K}, \delta_{40}= 0),
 \end{equation}
 which is the transition redshift at mean cosmic density for a typical critical virial temperature. This characterstic transition redshift is not used in the further analysis, and we merely include it to give readers an indication for the redshift at which the transition occurs in a typical case.
 
 We chose the basic functional form in Eq. \ref{Eq:func} because the $\arctan$ function turned out to be an excellent fit for the redshift dependence of \fII\ at fixed $\Tc$ and $\delta_{40}$ found in our simulations. We then added higher order terms in $T_3$ and $\delta_{40}$ until it was possible to fit \fII\  everywhere in the three-dimensional parameter space of redshift, $\Tc$ and $\delta_{40}$. The free parameters in this function are $F_0$, $A$, $\Delta z$, $a_1$, $a_2$, $a_3$ and $a_4$. The results of this fitting procedure are discussed below.

There are already several haloes with $T_\mathrm{vir} > 2000\,\mathrm{K}$ at the earliest time-step of the merger trees (around $z\approx 30$). This leads to an instantaneous burst of Pop~III formation in the first time-step for models with low \Tc\ and therefore to a jump in the Pop~II fraction exactly one recovery time later. This initial jump is a numerical artifact and we therefore exclude the time-steps smaller than one recovery time since the first snap-shot from our fits.

\begin{figure}
  \includegraphics[width=\linewidth]{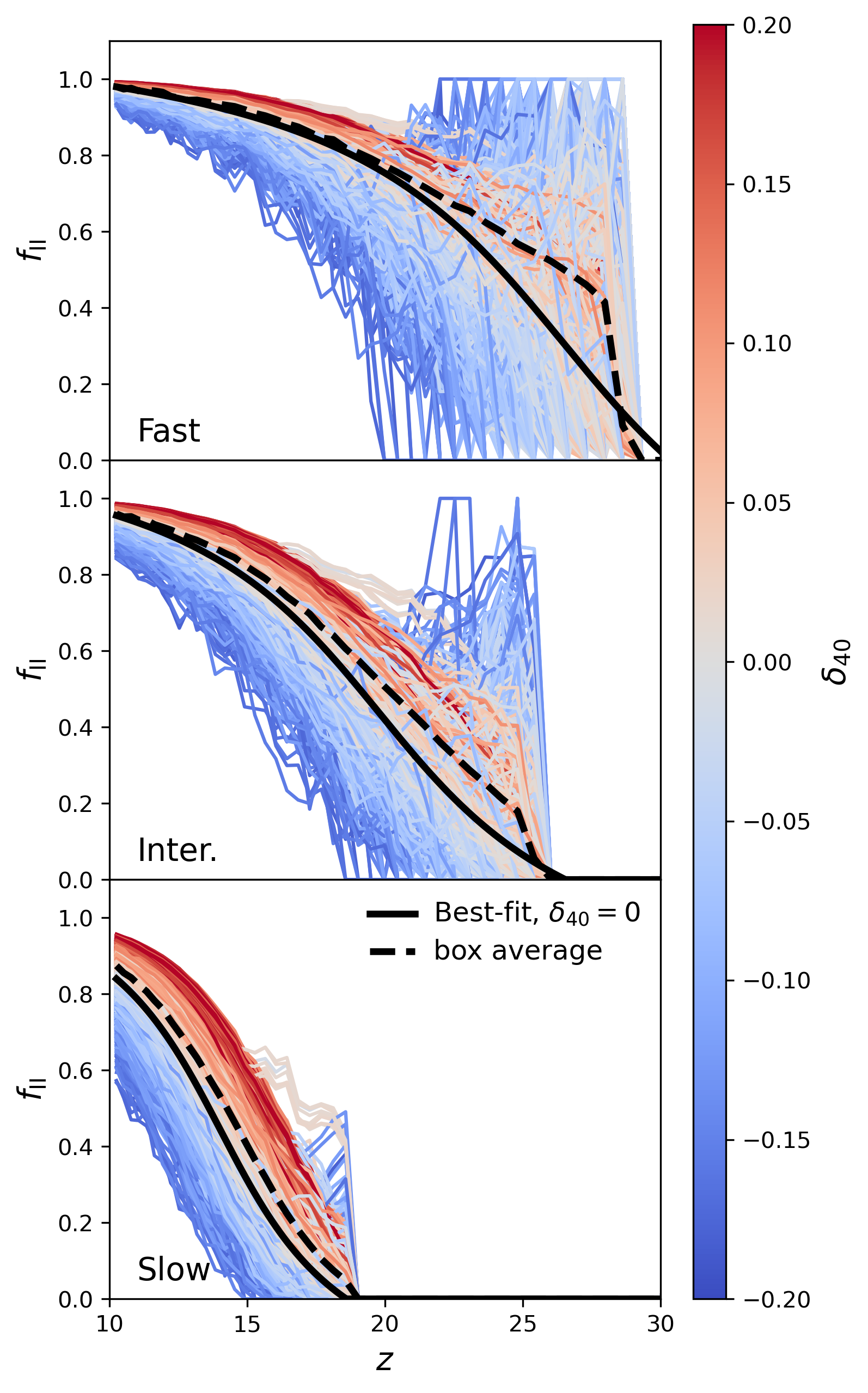}
  \caption{\label{fig:T2200} Example Pop~II fraction as a function of overdensity (colorbar) and redshift for a critical virial temperature of $\Tc=2200\,\K$ for all three recovery times: fast (top), intermediate (middle) and slow (bottom). We also show the Pop~II fraction of the entire simulated region (dashed black line) and the best-fitting curve at $\delta_{40}=0$ (black). Note that the average Pop~II fraction is not equal to the Pop~II fraction at average density. The noise in the low-density pixels at high redshift is due to low number statistics.
  }
 \end{figure}

We show an example of the Pop~II fraction for a critical virial temperature of $\Tc=2200\,\K$ in Fig. \ref{fig:T2200} for the three recovery times. We chose this particular virial temperature as an example because it has been found to be a suitable value for the first collapse of mini-haloes by \citet{Hummel12} and because around $z\approx20$ it results in a similar mass-threshold as found by \citet{Schauer19a} for moderate streaming velocities. Overall we can see that these fits represent the data well. As anticipated, the transition occurs earlier in regions with high overdensities. There is significant noise in the low-density pixels at high redshifts: as a result of small-number statistics, there is more noise in underdense regions, because they contain fewer haloes. As the low density pixels only contain a few haloes at these redshifts, the Pop~II fraction changes very significantly with each single halo that transits to metal-enriched star formation. We also find that quicker recovery times show more scatter because Pop~II star formation occurs at a time when there are fewer star-forming haloes overall. We present more detailed residuals of the fits in the space of $z$, $\delta$ and $\Tc$ in Appendix \ref{apx:res}.

The best-fitting parameters for the fast, the intermediate and the slow transition models are presented in Table \ref{tab:best_fit}. The characteristic transition redshift for $\Tc=2200\,\mathrm{K}$ ranges from $z_\mathrm{t}\approx 25$ in the fast case down to $z\approx 14$ in the slow case. We can also see that the transition at larger recovery times occurs over a shorter redshift interval $\Delta z$. However, this shortening of the transition period is only an effect of a similar redshift interval corresponding to more physical time at lower redshifts. The duration of the transition in physical time at the characterstic transition redshift $z_\mathrm{t}$ is $\Delta t\approx 30\,\mathrm{Myr}$ for the fast, $\Delta t\approx 72\,\mathrm{Myr}$ for the intermediate and $\Delta t\approx114\,\mathrm{Myr}$ for the slow transition.

As a consistency check, we estimate the redshift at which the transition from metal-free to metal-enriched star formation occurs in cosmological simulations from the First Billion Years project \citep[FiBY][]{FiBY1}, the Pop~III Legacy \citep[P3L][]{Jaacks19} and the \textit{Renaissance} simulations \citep{Xu16b,Xu16c}. The latter is a set of three simulations at different over-densities of $\delta_{40} = (0.27, 0.03, -0.06)$, which are referred to as `rarepeak', `normal' and `void'. For all these simulations, we integrate the star formation rates (SFRs) and find the redshift at which exactly 50 per cent of all the stars formed up to that time are metal-free. While this redshift is a marker of the transition to metal-enriched star formation, it is based on SFRs, rather than the halo properties. Thus, it does not have the same definition as our transition redshift defined in  Eq. \eqref{eq:z_0}. To emphasize this difference we refer to the redshift at which 50 per cent of stars formed are metal-free as `redshift of equality', rather than as transition redshift. Although the redshift of equality and the transition redshift have different definitions, they characterize the same physical process and, thus, are expected to be similar in value. We do not directly compute the transition redshift as we define it in Eq. \eqref{eq:z_0} from the simulations. This would require a detailed analysis that goes beyond the scope of this project. We also do not calculate the redshift of equality for the semi-analytical model, as it would introduce dependencies on the star formation efficiencies, which we prefer to avoid. The comparison is therefore qualitative in nature. The redshifts of equality are $z\approx14$ in FiBY, $z \approx 22.3$ in P3L and $z \approx 25$, 21.5, and 21 in \textit{Renaissance} for their `rarepeak', `normal' and `void' simulations respectively. In all these simulations, star formation predominantly occurs in haloes with a mass of $M_\mathrm{vir}\gtrsim 10^7\,\Ms$. At a transition redshift around $z\approx 16$ this corresponds to a critical virial temperature of $\Tc = 8000\,\K$ (i.e., $T_3=0.9$) and we, therefore, compare the redshifts of equality in the simulations with the fitted transition redshift for this critical virial temperature, i.e., $z_0(T_3=0.9\,\mathrm{K})$ according to eq. \eqref{eq:z_0}. We show the transition redshift for a range of different overdensities $\delta_{40}$.

\begin{figure}
  \includegraphics[width=\linewidth]{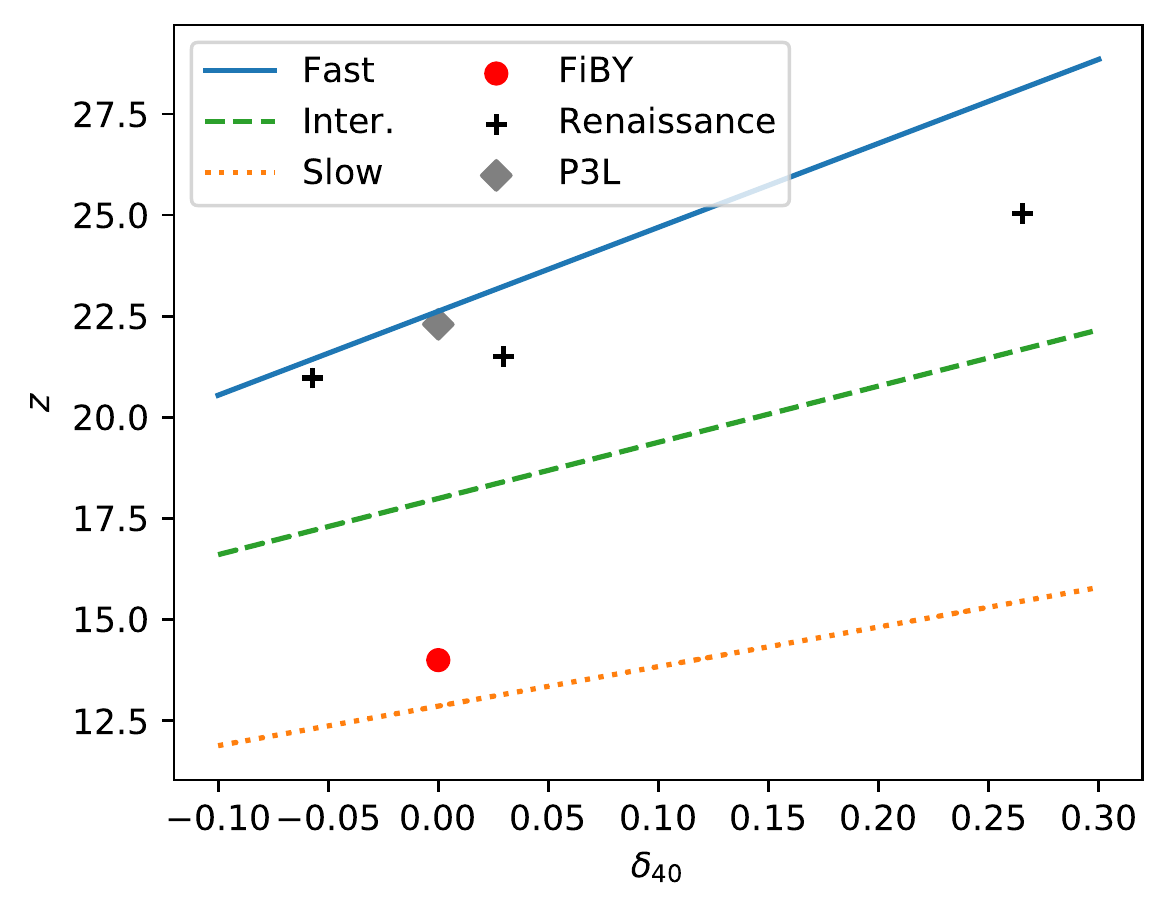}
  \caption{\label{fig:z_0}Qualitative comparison between our fitted transition redshift $z_0$ (defined in Eq. \ref{eq:z_0}, and calculated for $\Tc=8000\,\K$) and the redshifts of equality (see text) from FiBY, the Pop~III Legacy (P3L) and the \textit{Renaissance} simulations as a function of the large-scale overdensity at $z=40$, $\delta_{40}$. The comparison is qualitative owing to the discrepancy in the definitions of the transition redshifts.}
 \end{figure}

The comparison is presented in Fig. \ref{fig:z_0}. We find that, despite the large scatter, the redshifts of equality from the simulations are in a broad agreement with the predictions of our model. The transition redshift from FiBY is close to our $z_0$ in the case of the long recovery time (slow transition), while the ones from the P3L and \textit{Renaissance} simulations fall between the fast and the intermediate models. The change in redshift of equality as function of overdensity seen in the \textit{Renaissance} simulations is similar to our model predictions. We note that the simulations from all three mentioned projects have completely independent implementations and vary in their setup, assumptions and employed methods. It is not clear which of the assumptions cause the large variation in the redshifts of equality found in these simulations. However, the fact that the span of transition redshifts computed from our models covers the variety of the redshifts of equality found in these diverse simulations indicates that our approach brackets cases realized in state-of-the-art cosmological simulations.

 \begin{table}
 \caption{\label{tab:best_fit}Best-fitting parameters of the analytic expression of the Pop~II fraction. The fitting function is defined in Eq \eqref{eq:fitting_function}. We additionally show the characteristic transition redshift $z_\mathrm{t}$, which is $z_0$ (see Eq. \ref{eq:z_0}) at $\Tc=2200\,\mathrm{K}$ and $\delta_{40}=0$.}
 \begin{tabular}{l|lll}
 \hline
Model&Fast&Inter.&Slow\\
\hline
$t_\mathrm{recov}$ (Myr) & 10 &30&100\\
$F_0$&0.378&0.442& 0.452\\
$A$&0.484&0.481& 0.499\\
$\Delta z$&5.11&5.13& 3.62\\
$a_1$&3.74&2.33& 1.48\\
$a_2$&26&20.1& 14.2\\
$a_3$&-8.59&-3.98& -1.45\\
$a_4$&-13&-10.3& -8.47\\
$z_\mathrm{t}$ &24.7 & 19.3 & 13.7\\
 \end{tabular}
 \end{table}
 
 \section{Large-scale simulations of the 21-cm signal}
 \label{Sec:21cmSims}
The 21-cm brightness temperature is given by
\begin{equation}
    T_{\rm 21} = \frac{T_{\rm S} - T_{\rm CMB}}{1+z} \left( 1- e^{-\tau_{21}} \right),
\end{equation}
where $T_{\rm S}$ is the spin temperature (which corresponds to the excitation temperature of the hydrogen ground state), $\tau_{21}$ is the 21-cm optical depth \citep[which itself depends on the spin temperature,][]{madau97}, and $T_{\rm CMB} = 2.725(1+z)$ K is the temperature of the cosmic microwave background (CMB) radiation\footnote{In the presence of high redshift radio sources this temperature is replaced by the total radiation temperature at the wavelength of the 21~cm line \citep[][]{Feng:2018, reis20c}.}. The 21-cm signal can only be seen when the spin temperature is driven away from the background radiation temperature. During Cosmic Dawn and the EoR this is enabled through the subtle Wouthuysen and Field effect \citep[WF, ][]{wouthuysen52, field58}, in which the absorption and re-emission of \Lya{} photons by hydrogen atoms couple the spin temperature to the kinetic temperature of the gas. The source of these \Lya{} photons are stars in the first galaxies. After the coupling between the spin temperature and the gas temperature is established, the 21-cm signal is expected to be seen in absorption. This is expected because the gas temperature is thought to be lower than the background radiation temperature at this stage (since after thermal decoupling, the gas cooled faster than the radiation). Radiation from galaxies can also heat the gas \citep[via X-ray and \Lya{} heating, e.g.,][]{madau97,Chuzhoy:2007, Reis:2021}, potentially resulting in a 21-cm signal seen in emission, and ionize the gas leading to the disappearance of the 21-cm signal from the inter-galactic medium (IGM). The 21-cm signal is predicted to be non-uniform with fluctuations originating from several sources including non-uniform hydrogen density and velocity as well as fluctuating radiative backgrounds \citep[e.g.][]{visbal12, Fialkov14, cohen18, Reis:2021}. 

Although Cosmic Dawn is unconstrained by observations, it is generally thought that at the onset of star formation the 21-cm signal is dominated by the effects of \Lya{} physics as these photons are very efficient in coupling the gas. For instance, exploring a large set of 21-cm simulations with variable astrophysical parameters, \citet{cohen17, cohen18} found that the redshift of \Lya{} coupling can be anywhere between $z\sim 35$ and 15 depending on the properties of star-forming haloes. X-ray heating becomes relevant later \citep[$z\lesssim 20$, ][]{cohen17, cohen18} owing to the fact that it takes time for the first population of X-ray sources  \citep[such as X-ray binaries,][]{fragos13} to emerge. The impact of reionization on the 21-cm signal becomes apparent only at relatively late times \citep[$z\lesssim 15$ in][]{cohen17, cohen18} with the appearance of massive galaxies which are efficient in ionizing the gas \citep[e.g., ][]{Park:2020}. It is, thus, expected that the Pop~III-Pop~II transition explored in this paper will mostly affect the 21-cm signal from the era of \Lya{} coupling. Therefore, for simplicity, we will ignore the impact of X-ray and ionizing photons in this work. The only heating/cooling mechanisms that might affect the temperature of the IGM in the absence of X-ray and ionizing photons are cooling due to the expansion of the Universe, heating by the \Lya{} photons \citep{Chen2004, Chuzhoy:2007, Reis:2021} and the CMB \citep{venumadhav18}, as well as the impact of structure formation. All of these effects are included in the simulation.

We use our own semi-numerical code to calculate the 21-cm signal \citep[e.g.,][]{visbal12, Fialkov14, cohen17, fialkov19, reis20a, Reis:2021}. The simulation size is 384$^3$ Mpc$^3$ and resolution is 3 comoving Mpc. The outputs of the simulation are cubes of the 21-cm brightness temperature at each redshift. From these we calculate the global signal and the spherically averaged power spectrum. The input of the simulation is a realization of the initial density and velocity fields \citep[calculated using publicly available code \texttt{CAMB},][]{camb}. The density and velocity fields are then evolved using linear perturbation theory. To calculate the population of dark matter haloes given the density field we use the hybrid approach of \citet{barkana04} which combines the previous models of \citet{press74} and \citet{sheth99}. In the simulation, the minimum halo mass for star formation is parametrized by the circular velocity $V_c$ which is related to $\Tc$ by
\begin{equation}
\Tc = \frac{\mu m_p V_c^2}{2 k_{\rm B}} = 7300\,\mathrm{K} \left(\frac{V_c}{10\,\mathrm{km}\,\mathrm{s}^{-1}} \right)^2, 
\label{eq:Tcrit}
\end{equation}
where $\mu$ is the mean molecular weight, $m_p$ is the proton mass, and $k_{\rm B}$ is the Boltzmann constant. Note that the value of $\mu$ depends on the ionization fraction of the gas. Here, we assume neutral primordial gas for which $\mu = 1.22$. Our simulation includes the effect of the relative velocity between dark matter and gas \citep[following the prescription in][]{Fialkov12,visbal12}, Lyman-Werner feedback \citep[as described in][]{fialkov2013} and photoheating feedback \citep[from][]{cohen16} on the minimum halo mass for star formation. 

Haloes accrete gas and convert it into stars with star formation efficiency which is constant (denoted by $f_*$) for halo masses above the atomic cooling threshold but drops as the logarithm of mass at lower masses \citep{cohen17}. To relate the stellar mass to the radiation produced in the \Lya{} line and the LW bands we use our fiducial model for Pop~III and Pop~II emissivities based on the results of \citet{barkana05}. Our Pop~II model is calibrated to the locally measured Scalo IMF \citep{Scalo:1998} with a metallicity of 5\% of the solar value. Pop~III stars are all assumed to be 100 solar mass, which was the prediction of the early Pop~III simulations by \citet{Abel02}. Stellar spectra are approximated by a series of power law curves (a separate power law for every pair of consecutive levels of atomic hydrogen).

 \subsection{Implementing the Pop~III - Pop~II transition in 21-cm simulations}
\label{sec:transition_impl}
 
Previously, in our 21-cm simulations Pop~III and Pop~II stars were implemented on equal footing, with the total mass in stars calculated using the accreted gas mass and assuming certain star formation efficiency \citep[e.g. see][ for details]{cohen20}. The only difference between Pop~III and Pop~II stemmed from the different stellar emissivities. Only one type of stars (either Pop~III or Pop~II) could be selected. In this work we differentiate between Pop~III and Pop~II star-forming haloes using the results of Section \ref{sect:fitting} and change the prescription for the calculation of the stellar mass in Pop~III forming haloes (following the single-burst approximation). As before, each type of stars produces radiation according to their respective emissivities. The contributions of Pop~III and Pop~II to the evolving and fluctuating radiative backgrounds that drive the 21-cm signal are computed separately and then added up. 

\begin{figure}
    \centering
    \includegraphics[width=\linewidth]{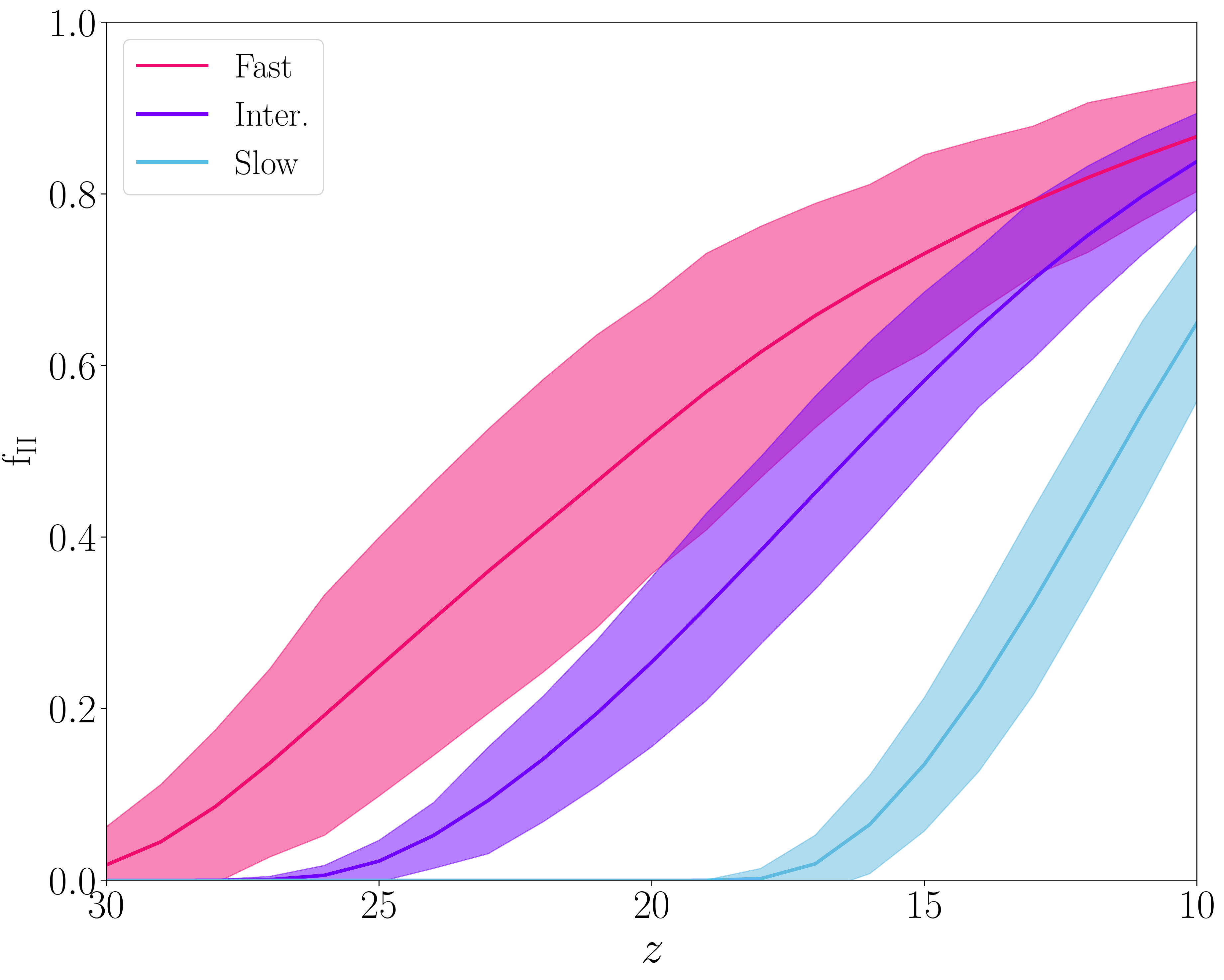}
      \includegraphics[width=\linewidth]{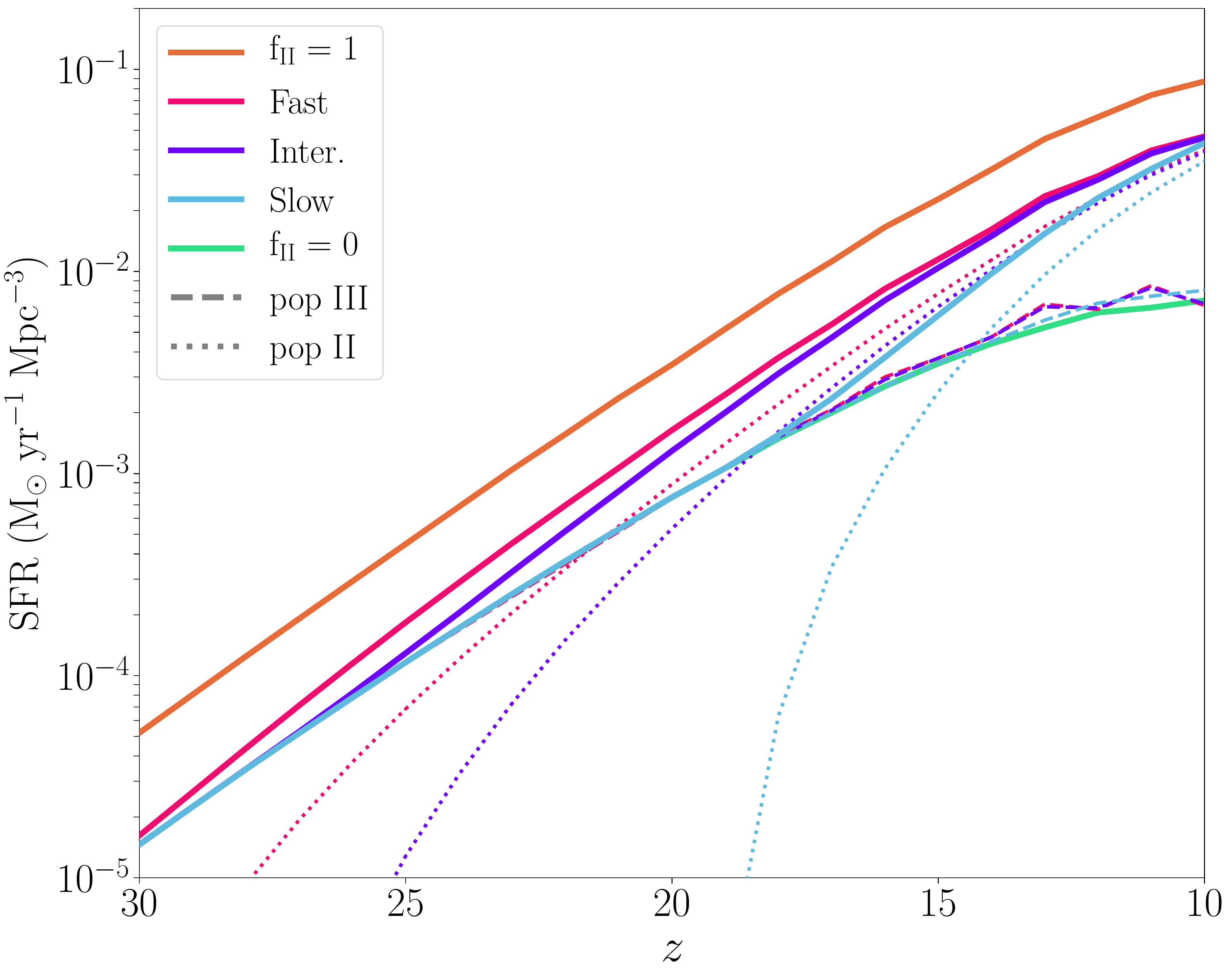}
    \caption{{\bf Top:} The mean Pop~II fraction $f_{\rm II}$ (solid lines) calculated from the 21-cm simulations shown for $V_c$ = 5.4 km s$^{-1}$ for the three time-delays: fast (magenta), intermediate (purple), and slow (cyan). The shaded regions show the corresponding standard deviation in $f_{\rm II}$  calculated over one simulation box of 384$^3$ Mpc$^3$ at the resolution corresponding to the pixel size of 3$^3$ Mpc$^3$. {\bf Bottom:} The SFR shown for $V_c$ = 5.4 km s$^{-1}$, and $f_{*, \rm III}=f_{*, \rm II} = 0.05$ and the three cases of star formation transition: fast (magenta), intermediate (purple), and slow (blue). We show the total result (solid) as well as the individual contributions of Pop~III (dashed) and Pop~II (dotted) stars. We also show the case with Pop~III stars only (denoted as  $f_{\rm II}=0$, turquoise) and the full Pop~II stars case ($f_{\rm II}=1$, orange). 
    }
    \label{fig:fII}
\end{figure}

The total Pop~II stellar mass in each pixel of the 21-cm simulation box is computed by multiplying the total mass in stars, calculated as previously using the accreted gas mass and assuming a star formation efficiency $f_{*,\rm II}$, by the fraction of haloes that actually form Pop~II stars, $f_{\rm II}$. Because the Pop~II fraction is a function of redshift, $\Tc$ (related to $V_c$ by Eq. \ref{eq:Tcrit} and affected by the non-uniform LW feedback and streaming velocities) and local overdensity, the Pop~II content is inhomogeneous and $f_{\rm II}$ varies across the simulation box. Examples of the mean value and scatter in $f_{\rm II}$ calculated from the 21-cm simulations are shown in the top panel of Fig. \ref{fig:fII} for an astrophysical scenario with $V_c$ = 5.4 km s$^{-1}$ (corresponding to $\Tc =2200\,\mathrm{K}$) and for the three cases of Pop~III-Pop~II transition (fast, intermediate, and slow).

To calculate the contribution of Pop~III star forming haloes to star formation in every pixel and at a given redshift we find the number of haloes above the star formation threshold ($M_\mathrm{crit}$) that have formed within an interval of time equal to the lifetime of Pop~III stars, $t^{\rm pop III}_{\rm lifetime}$. We then assume that each such halo produces 
\begin{equation}
M_{*,\rm III} = f_{*,\rm III} \frac{\Omega_\mathrm{b}}{\Omega_\mathrm{m}} M_\mathrm{crit}
\label{eq:PopIIIstars}
\end{equation}
 of Pop~III stars (in agreement with Eq. \ref{eq:MpopIII}). Note that we cannot simply use $1 - f_{\rm II}$ to calculate the Pop~III contribution since $1 - f_{\rm II}$ also includes inactive haloes that have already stopped forming Pop~III stars but have not yet started forming Pop~II. We also note that for the same values of $f_{*,\rm II}$ and $f_{*,\rm III}$, Pop~III formation is a less efficient process compared to the formation of Pop~II stars. This is because a halo of mass $M$ forming Pop~III stars will generate total stellar mass given by Eq. \ref{eq:PopIIIstars}, which is smaller than the stellar mass in Pop~II stars, $f_{*,\rm II}\left(\Omega_\mathrm{b}/\Omega_\mathrm{m}\right)M$, that the same halo can host. 
 
The bottom panel of Fig. \ref{fig:fII} shows the box-averaged contributions of Pop~III and Pop~II stars to the total SFR for slow, intermediate and fast transition. Here for simplicity we assume $f_{*,\rm II} = f_{*,\rm III} = 0.05$. In addition, we show the cases with Pop~III stars only (referred to as  $f_{\rm II}=0$) and the full Pop~II stars case ($f_{\rm II}=1$) where we assume that Pop~II stars form from the start, with no recovery delay, and there is no episode of Pop~III star formation.

We see that the contribution from Pop~III to the total SFR is the similar (up to a small discrepancy explained by the difference in $ M_\mathrm{crit}$, which is a result of the variation in the LW feedback driven by the  difference in the total SFR) in all transition scenarios. The scenarios differ in the contribution from Pop~II: Owing to the rapid early rise in the number of Pop~II star-forming haloes in the case of the intermediate and fast transitions, the total SFR is driven by the metal-enriched population through the most part of cosmic history. On the contrary, in the slow transition case primordial stars dominate SFR (and, thus, will drive the 21-cm signal, as we discuss later) all the way down to $z\sim 17$. At lower redshifts, the rapid increase in the number of Pop~II star forming haloes in this case results in a fast growth of SFR. 

\begin{figure}
     \includegraphics[width=\linewidth]{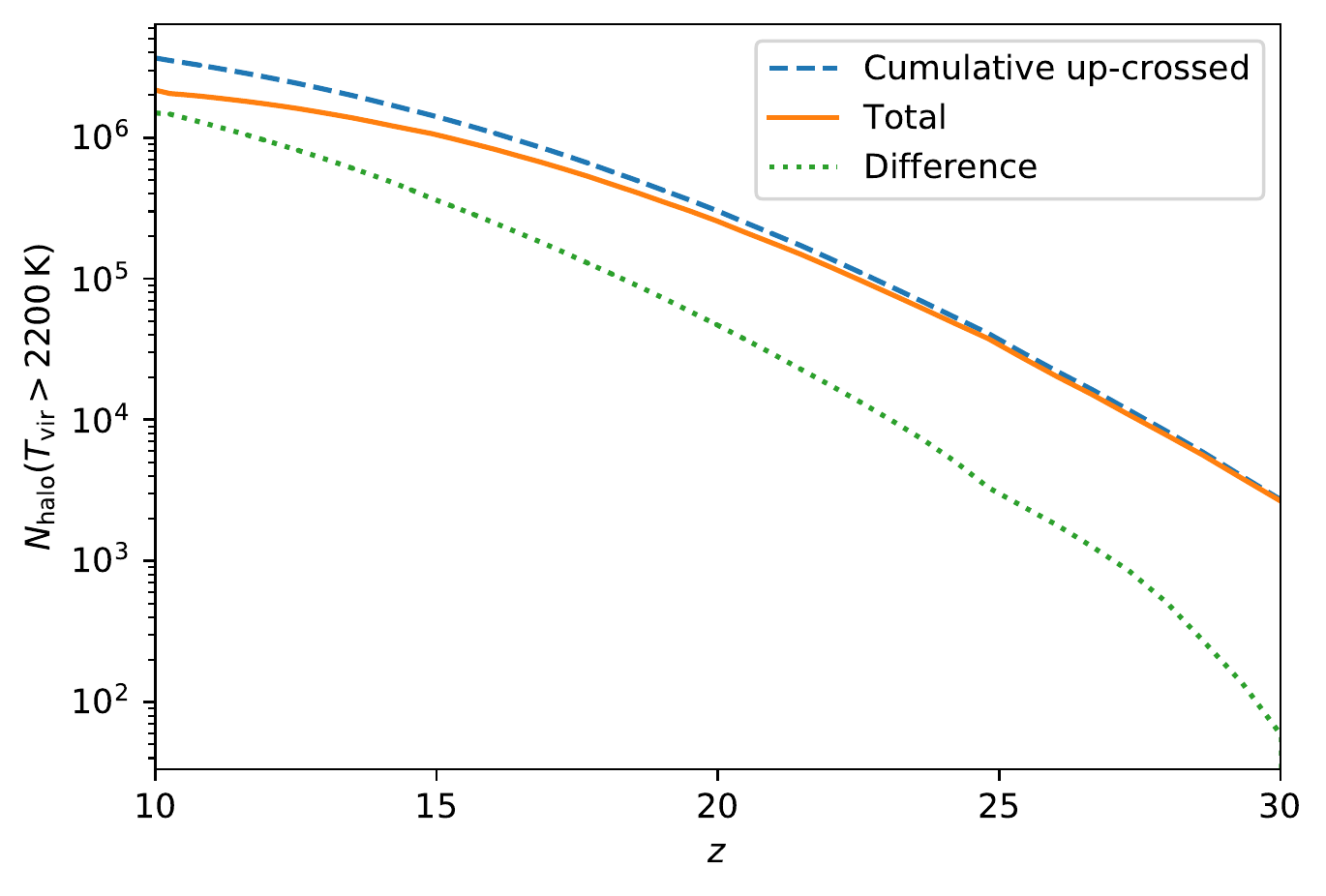}
     \caption{Total number of haloes above a critical virial temperature of $\Tc=2200\,\K$ as a function of redshift (orange solid line) compared to the total number that ever exceeded $\Tc$ until that redshift (blue solid line) and the difference between the two (green). If there were no mergers of haloes above $M_\mathrm{crit}$ the two lines should be the same and the difference should be zero.}
     \label{fig:upcross}
 \end{figure}

Finally, we note that the implementation of this model relies on the assumption that the total number of haloes above the critical mass is a good approximation for the total number of haloes that ever formed Pop~III stars. This is justified if the number of haloes that ever crossed the critical mass threshold is similar to the number of haloes that are above the critical mass threshold, i.e., that haloes mostly grow via smooth accretion, while mergers between haloes above the critical mass are rare. To ensure that this assumption is fulfilled we compare the total number of haloes above $\Tc=2200\,\K$ as a function of redshift to the total number of haloes that ever reached $\Tc=2200\,\K$ until that redshift using the results of the $N$-body simulations from Section \ref{sect:nbody} (see Fig. \ref{fig:upcross}). Indeed, we can see that mergers do not affect the halo numbers above \Tc\ strongly for $z>15$, where our Pop~III star formation model is most important. At lower redshifts the expected 21-cm signal is dominated by Pop~II star formation, which is only dependent on the halo mass function and not on their merger histories. However, we note that the discrepancy reaches about a factor of two at redshift $z=11$, indicating that mergers should not be neglected at lower redshifts.

 \section{Effect of Pop~III - Pop~II transition on the 21-cm signal from Cosmic Dawn}
 \label{sec:res}

In the absence of X-ray heating sources, the 21-cm signal from Cosmic Dawn is largely driven by \Lya{} photons and, therefore, is tightly linked to the SFR. In this section we explore the impact of the Pop~III-Pop~II transition on the characteristic features of the 21-cm signal, including the typical deep absorption trough in the global signal and the peak in the power spectrum imprinted by the inhomogeneous \Lya{} coupling and \Lya{} heating. 

\begin{figure}
    \centering
    \includegraphics[width=\linewidth]{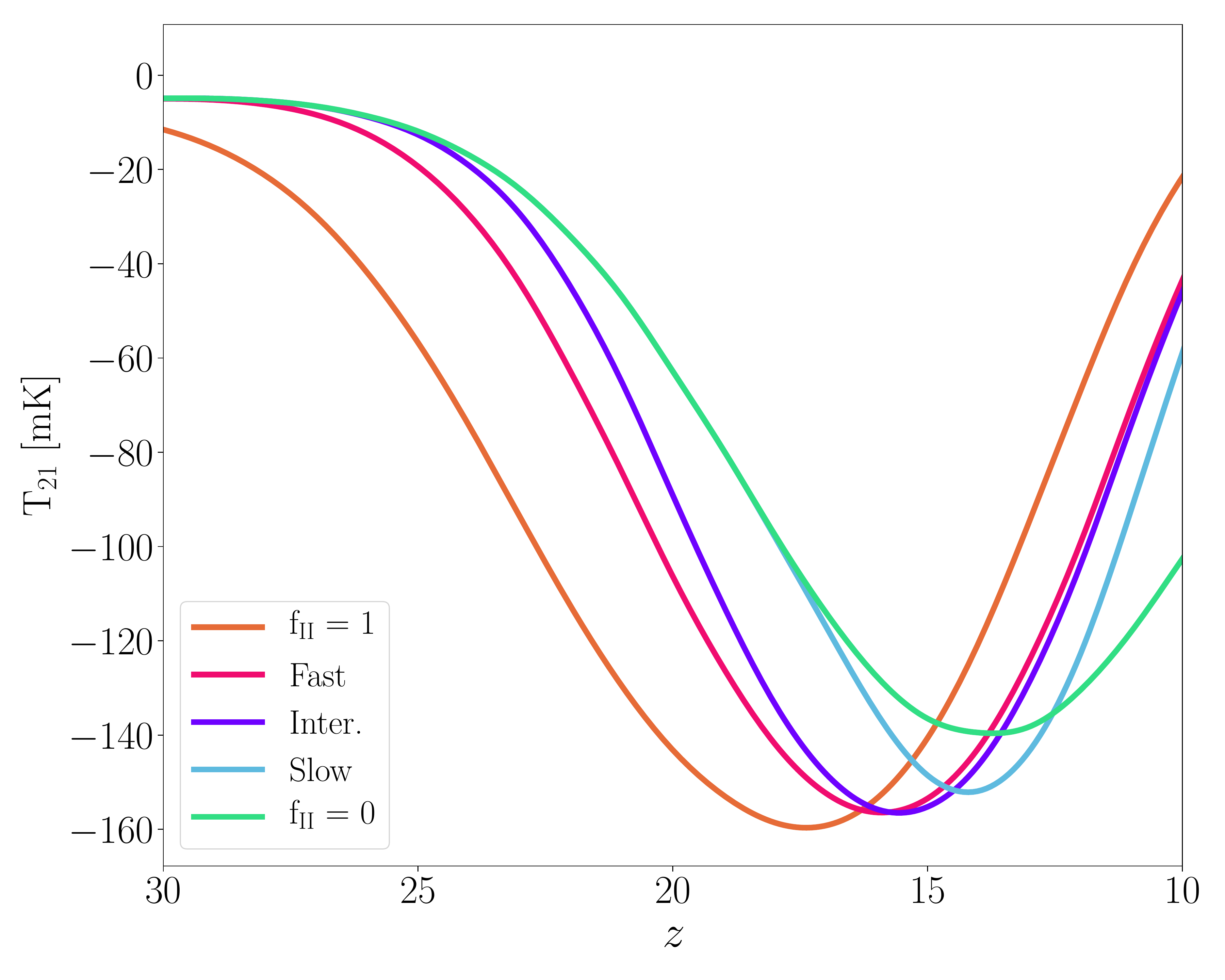}
    \includegraphics[width=\linewidth]{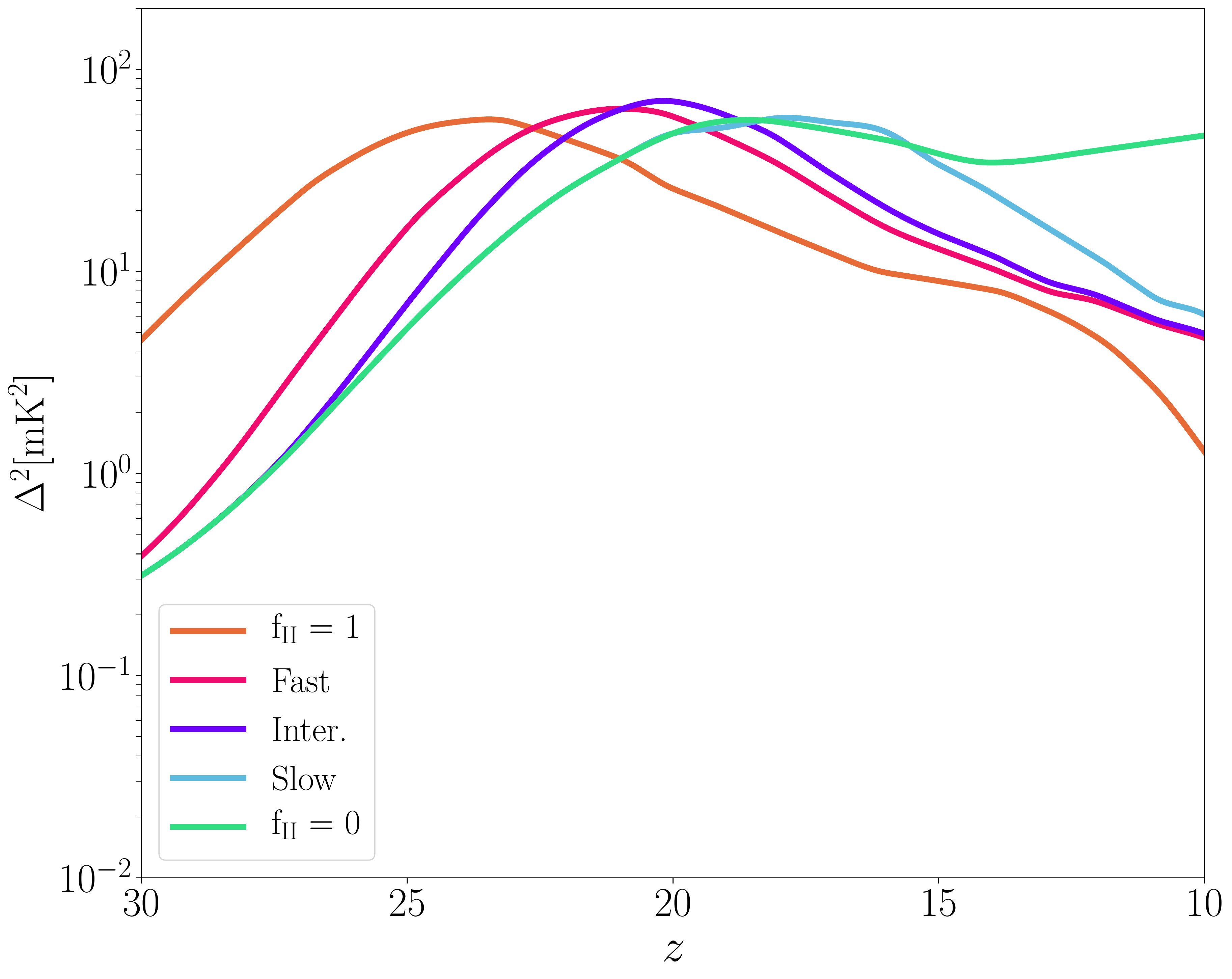}
    \caption{Examples of the effect of the transition to Pop II star formation on the 21-cm signal. {\bf Top:} The global signal. {\bf Bottom:} The power spectrum at k = 0.1 Mpc$^{-1}$. We show a model with $V_c$ = 5.4 kms$^{-1}$ and $f_{*, \rm III}=f_{*, \rm II} = 0.05$ for the three scenarios with fast (magenta), intermediate (purple), and slow (blue) transition. We also show two reference models, one with $f_{\rm II} = 1$ (orange) and one with $f_{\rm II} = 0$ (turquoise).}
    \label{fig:21cm_signal}
\end{figure}

Fig. \ref{fig:21cm_signal} shows the redshift dependence of both the global 21-cm signal and its power spectrum calculated for the same astrophysical scenarios that were used to demonstrate the effect of the population transition on the SFR (shown in Fig. \ref{fig:fII}). In addition, here we plot the two limiting cases, $f_{\rm II} = 1$ (full Pop~II case) and $f_{\rm II} = 0$ (only Pop~III stars). Echoing the growth of SFR with time, the more realistic 21-cm signals which include the Pop~III-Pop~II transition evolve faster compared to the reference case with Pop~III stars only. The higher SFR of the realistic models is manifested by the shift to higher redshifts of the absorption feature in the global signal and the earlier emergence of the corresponding peak in the power spectrum. The resulting global signal is stronger with a deeper and narrower absorption trough (the former is due to the more efficient \Lya{} coupling and the latter is a manifestation of the more efficient \Lya{} heating compared to the Pop~III-only case). Correspondingly, the peak in the power spectrum is higher (although this is a small effect) and narrower. On the other hand, compared to the $f_{\rm II} = 1$ case, the realistic scenarios, which inevitably include an initial phase of Pop~III star formation, evolve slower.

In addition to the general impact of the SFR on the timing of the signal, the character of the transition between the two stellar populations is reflected in the shape of the 21-cm signal. In all the realistic scenarios the initial stage of the process of \Lya{} coupling is dominated by Pop~III star formation, as is evident from the overlapping high-redshift parts of the two 21-cm signals corresponding to the scenario with $f_{\rm II} = 0$ and the slow transition model. However, the onset of Pop~II star formation, accompanied by a boost in the number of \Lya{} photons, leads to a divergence of these two signals with a Pop III-only case lagging behind. The rapid growth of the enriched population results in a steepening of the 21-cm signal (seen both in the global signal and the power spectrum), which is a potentially testable prediction. Similar, but much stronger effect of the emerging Pop~II formation is seen in the signals corresponding to the intermediate and fast transitions which deviate from the $f_{\rm II} = 0$ case very early on and are steeper than both $f_{\rm II} = 0$ and $f_{\rm II} = 1$ cases. In these scenarios the contribution of Pop~II stars is important throughout the Cosmic Dawn and affects the 21-cm signal over a broad range of redhifts.

\begin{figure*}
    \centering
    \includegraphics[width=0.45\textwidth]{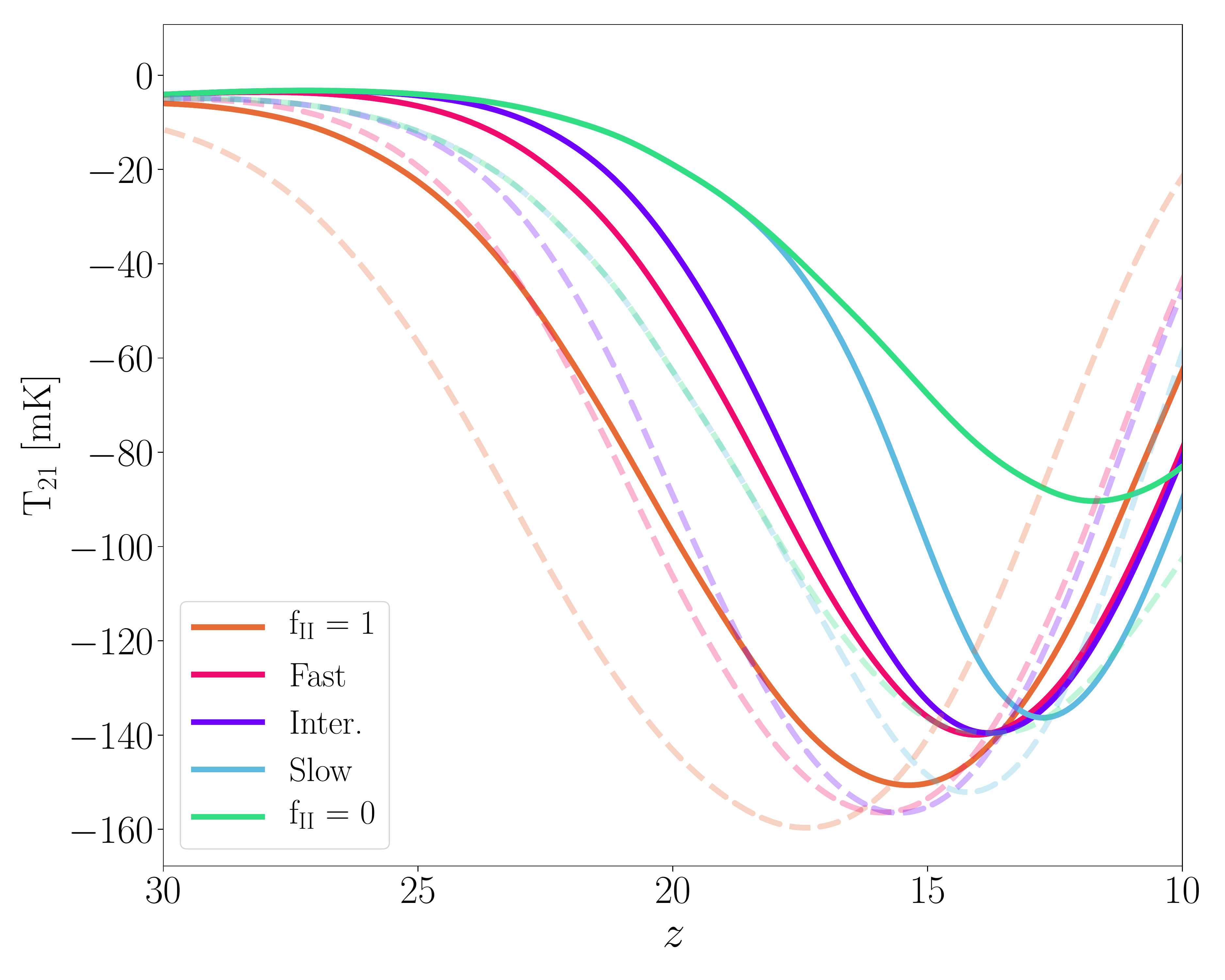}
    \includegraphics[width=0.45\textwidth]{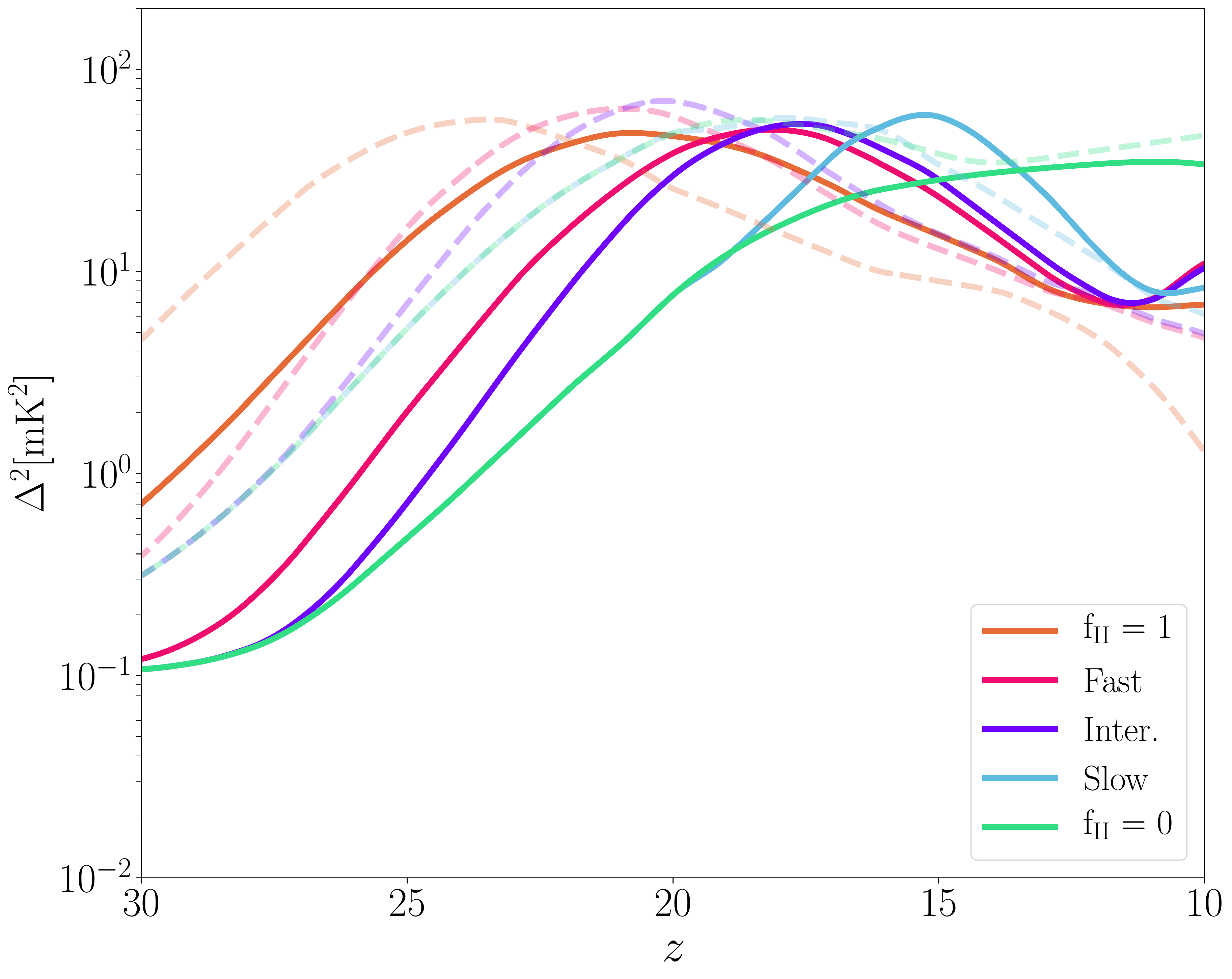}
    \includegraphics[width=0.45\textwidth]{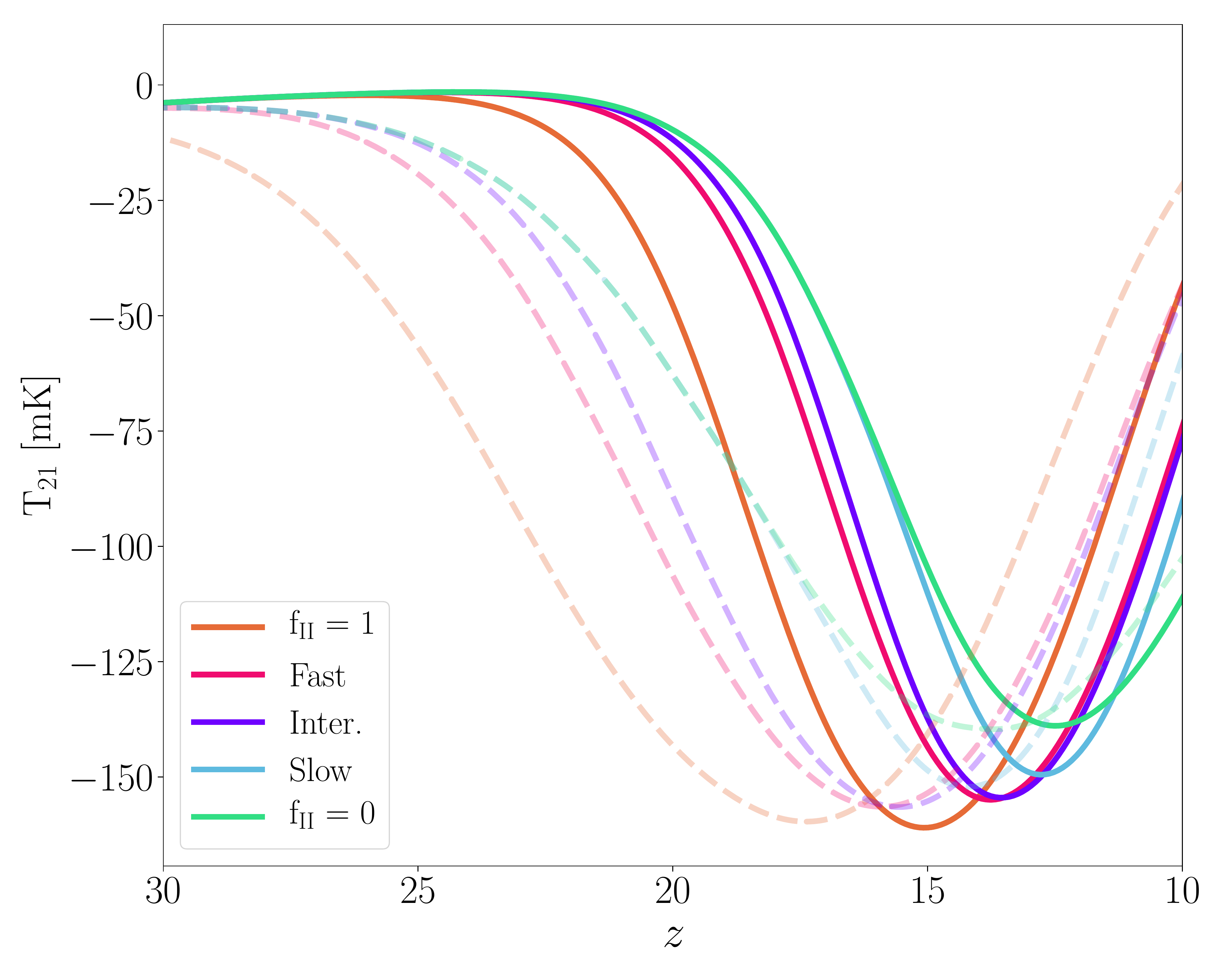}
    \includegraphics[width=0.45\textwidth]{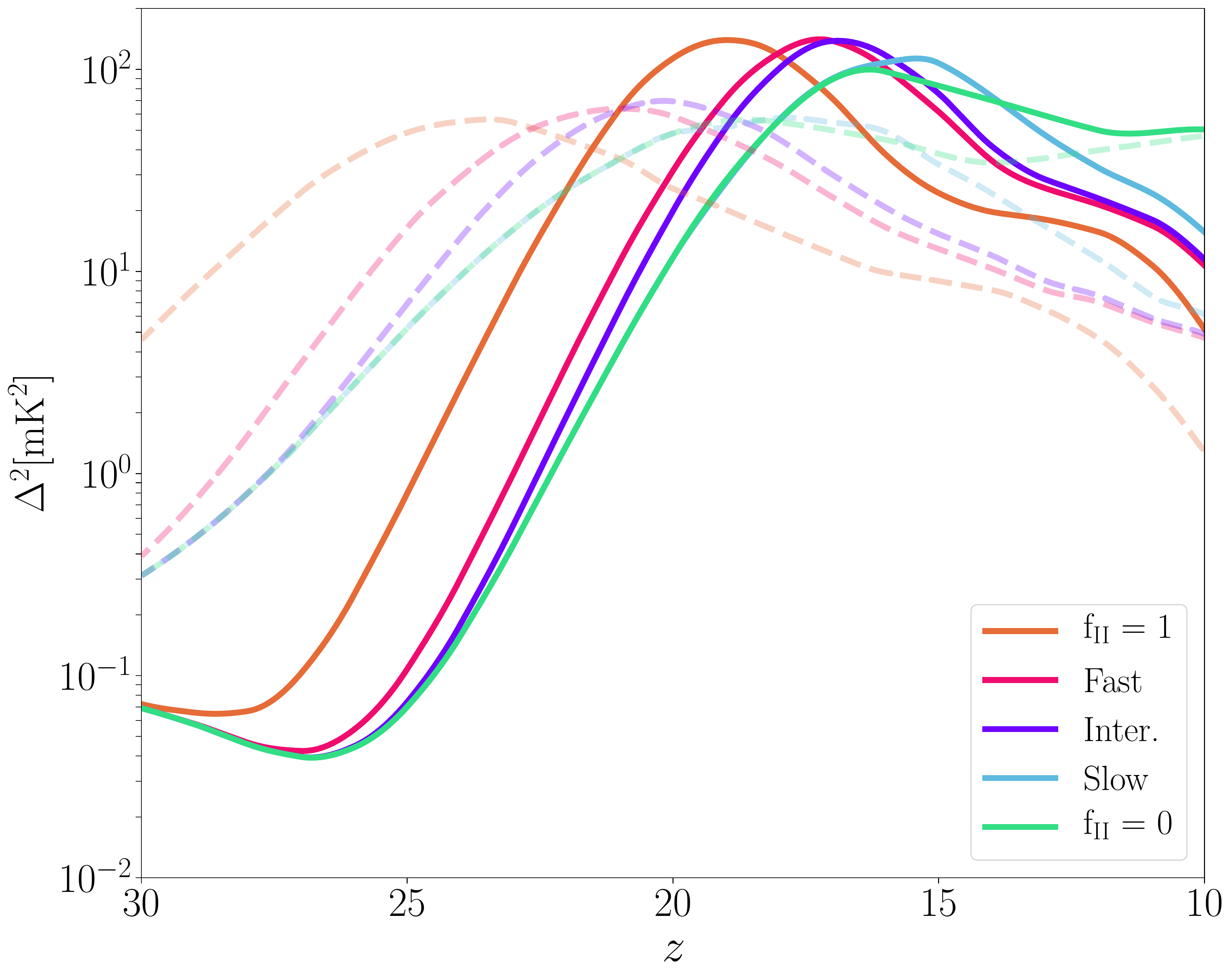}
    \caption{Global 21-cm signal (left) and the power spectrum (right) for different astrophysical parameters. Top: Same as Fig. \ref{fig:21cm_signal} but for $f_{*, \rm III} = f_{*, \rm II} = 0.01$. Bottom: $V_c$ of 35.5 km s$^{-1}$ (corresponding to $\Tc = 9.3\times 10^4$ K) and $f_{*, \rm III} = f_{*, \rm II} = 0.1$. For comparison, we also show the curves from Fig. \ref{fig:21cm_signal} (dashed). }
    \label{fig:21cm_signal_low_fstar}
\end{figure*}

The signature of the Pop~III-Pop~II transition on the 21-cm signal is model-dependent and varies as a function of astrophysical parameters, as can be seen from the two additional cases shown in Fig. \ref{fig:21cm_signal_low_fstar}. The first scenario (top panels of Fig. \ref{fig:21cm_signal_low_fstar}) has the same critical temperature as our main model (from Fig. \ref{fig:21cm_signal}, also shown with faint dashed curves in Fig. \ref{fig:21cm_signal_low_fstar} for comparison), but less efficient star formation with $f_{*, \rm III} =f_{*, \rm II}=0.01$ compared to 0.05 in the main case. Because the redshift evolution of the number of Pop~II forming haloes (determined by $f_{\rm II}$) is independent of star formation efficiency, features of the 21-cm signal that depend on the properties of the Pop~III-Pop~II transition (such as the relative steepness of the signals and the redshift at which the slow transition curve diverges from the $f_{\rm II} = 0$ reference case) are the same as in our main case. On the other hand, the absolute strength of the Cosmic Dawn signal is mostly determined by the intensity of the \Lya{} background and, therefore, directly depends on the values of $f_{*, \rm III}$ and $f_{*, \rm II}$. One major difference between the cases with low and high star formation efficiency is that in the former case the 21-cm signal evolves slower, and so when it peaks there are more Pop~II forming haloes compared to the latter case. This example shows that for a fixed value of $V_c$ the contribution of Pop~III stars is more important in high $f_*$ models, where major milestones in the evolution of the 21-cm signal occur at higher redshifts where there are more Pop~III star forming haloes.

The final example that we consider here has a higher $\Tc = 9.3\times 10^4$ K (corresponding to $V_c$ of 35.5 km s$^{-1}$) compared to our main setup as well as a higher star formation efficiency $f_{*, \rm III} = f_{*, \rm II} = 0.1$ (bottom panels of Fig. \ref{fig:21cm_signal_low_fstar}). This model has a star formation threshold far above the atomic cooling limit, and is only considered as an extreme case. Even though the intensity of the global 21-cm signal is roughly the same in these models, the Pop~III-Pop~II transition happens much later (shifted by $\delta z$ between two and six) for the models with $V_c=35.5$ km s$^{-1}$. As a result, the related features such as the redshift at which the slow transition curve diverges from the $f_{\rm II} = 0$ reference case, are shifted to lower redshifts.

The approach presented in this paper provides a flexible basis to test the nature of the Pop~III-Pop~II transition using 21-cm observations. Because $f_{\rm II}$ is independent on star formation efficiency, it might be possible to measure the time-delay while marginalizing over the rest of the model parameters \citep[in a similar way limits on astrophysical parameters were calculated, e.g. using early data of EDGES and LOFAR by][]{Monsalve:2019,Mondal:2020}. However, such analysis is out of the scope of this paper and we leave it for future work.

\section{Discussion}
\label{sec:disc}
\subsection{Comparison to earlier works}
Properties of Pop~III stars have previously been shown to have an impact on the 21-cm signal \citep{cohen16, Mirocha18, Mebane18, T_Tanaka18, T_Tanaka21, Schauer19b, Mebane20}. In these models the biggest factors in determining the 21-cm background are the star formation efficiencies and the IMF averaged radiation output of the first stars. These previous studies focus only on Pop~III stars or introduce a very simple model for the transition. Here we expand this picture by quantifying the transition from Pop~III to Pop~II star formation based on a semi-analytical model motivated by numerical simulations. We show that the recovery time, which is determined by the efficacy of Pop~III stellar feedback, has a distinctive imprint in the 21-cm background, affecting both the global signal and the fluctuations. We have seen in Section \ref{sect:fitting} that the range of characteristic redshifts at which the transition to Pop~II star formation occurs in our model agrees with the results of cosmological simulations. A better understanding of how the recovery time arises will be key to interpreting upcoming 21-cm observations.

Here we briefly compare our model and results to the work of \citet{Mirocha18}, based on the method of \citet{Mebane18}, where the impact of Pop~III-Pop~II transition on the global 21-cm signal was investigated\footnote{To our knowledge, ours is the first work to explore the impact of the transition on the fluctuations of the 21-cm signal.}. The modelling of the transition presented here significantly differs from the one adopted by \citet{Mirocha18}. Most importantly, we assume that there is only one episode of Pop~III star formation per halo, and that first supernovae eject metal-enriched gas which re-collapses after a recovery time initiating the process of Pop~II star formation. In contrast, the recovery times are not implemented directly in the prescription of \citet{Mebane18}, where two modes for the Pop~III-Pop~II transition are considered: an energy-based and a momentum-based prescription. These prescriptions compare the energy (or momentum) injected into a halo by the SNe to the energy (or momentum) required to remove gas from the halo, and expel gas and metals accordingly. In particular, if the halo is small enough all gas and metals will be removed. The halo then proceeds to accrete pristine gas. The transition to Pop~II formation occurs if the gas-metallicity within the halo exceeds a certain threshold. Thus the transition is governed by the halo's ability to retain SN ejecta which is computed from the escape velocity (or binding energy) of the halo. This assumption results in a rapid succession of up to $\mathcal{O}(10)$ episodes of Pop~III star formation in the same halo. To our knowledge, this behaviour is not reflected in hydrodynamical simulations of Pop~III SNe \citep{Ritter12, Ritter15, Ritter16, Jeon14, bsmith15, Chiaki16}. In these simulations minihaloes retain enough metals to form Pop~II stars shortly after the first supernovae explosions. The limiting factor is that very energetic SNe may evacuate minihaloes to such a degree that a next episode of star formation could be delayed by up to or above 100\,Myr \citep[][corresponding to our longest recovery time]{Whalen08b}. However, we are unaware of any hydrodynamical simulations in which the stars forming in such a context would be metal-free.

We find that the redshift of the Pop~III to Pop~II transition from their energy-based model matches well with our results. The momentum-based model, which is also used for further predictions by \citet{Mebane20}, produces much higher transition redshifts ($z_\mathrm{t}>30$) than what is found in hydrodynamical simulations \citep[$13<z_\mathrm{t}<25$][]{FiBY1, Xu16b, Xu16c, Jaacks19}. 
 
\subsection{Measuring the recovery time with the 21-cm signal}
As we have shown in this work, the slow Pop~III-Pop~II transition mode leads to a 21-cm signal that is very different from either the intermediate or the fast transitions. In this mode, the early onset of the signal is dominated by Pop~III stars for a significant part of cosmic history, which is evident from the corresponding Cosmic Dawn signal that closely follows the Pop~III-only case. This slow transition, with a recovery time of 100\,Myr is associated with massive (above 100\,\Ms) Pop~III stars and their very energetic pair-instability SNe \citep{Jeon14, Chiaki16} or with many SNe in the same halo \citep{Ritter15}. In such models the \Lya{} coupling and heating happen later, typical 21-cm signals are shifted to lower redshifts (higher frequencies) compared to the models with fast/intermediate transition, and the late onset of Pop~II formation leads to a characteristic steepening of the signal. This dependence on the recovery time could allow us to indirectly measure this parameter as well as constrain the primordial IMF and star formation efficiency using the 21-cm data from either radiometers or interferometers.  We note that the Pop~III IMF also affects the radiation output of the first stars, both in terms of total emission power and in terms of the spectral energy distribution. Investigating these effects on the 21-cm signal will be subject to a follow-up study (Gessey-Jones et al. in prep.).

If the EDGES detection is confirmed to be a cosmological 21-cm signal, its timing at $z\sim17$ \citep{EDGES18} implies early star formation. Models consistent with this signal are characterized by efficient emission of \Lya{} photons as well as a strong X-ray background at high redshifts \citep[e.g.][]{Schauer19b, Fialkov18, fialkov19, Mirocha:2019, reis20c}, which is unlikely in the case of the slow Pop~III-Pop~II transition that we considered here, but could be easily achieved in the scenarios with either a fast or an intermediate transition. This would in turn indicate that only a single or very few SNe per halo took place, and that they had relatively low explosion energies.

\subsection{Future work}
As the transition from Pop~III to Pop~II star formation is predicted to happen at the high redshifts of Cosmic Dawn, we focus here on the signature of sources emitting radiation in \Lya{} and LW bands. While the evolving metallicity will also affect the luminosity of X-ray binaries formed as the first population of stars dies \citep{fragos13}, we do not consider this effect here, leaving the self-consistent modelling of the X-ray signature to future work. The EoR is predominantly driven by later-time evolved galaxies which are expected to be metal-rich and, therefore, we do not expect the Pop~III-Pop~II transition to have an important effect on reionization apart from a minor effect on the high-redhsift tail that can be constrained using the CMB polarization data \citep[e.g.][]{Heinrich2018}. With the consistent inclusion of the X-ray and UV sources, we will be able to constrain the delay-time in the formation of Pop~II stars from data.

\section{SUMMARY}
\label{sec:conc}
In this work we considered for the first time the effect of the transition from primordial star formation (Pop~III) to the first generation of metal-enriched stars (Pop~II) on the inhomogeneous 21-cm signal from Cosmic Dawn. Stars directly affect the 21-cm signal by emitting ultraviolet radiation and, therefore, the change in the mode of star formation will be imprinted in the shape of the 21-cm signal. Because the duration and timing of this transition is linked to the stellar IMF and the typical mass of the first star-forming haloes, the 21-cm signal from Cosmic Dawn can be used to constrain these properties.

We model the transition using the semi-analytical code \textsc{a-sloth} and compare signatures of fast, intermediate and slow transitions (with recovery times of 10, 30 and 100\,Myr respectively) in the 21-cm signal. We find that the fast and intermediate transitions, linked to low efficiency of Pop~III formation, weak feedback and a quick recovery after the first SNe, lead to a steeper 21-cm signal compared to all the rest of the explored scenarios. Such models are more likely to explain the tentative 21-cm signal reported by the EDGES collaboration, compared to the slow Pop~III-Pop~II transition characteristic of the case in which stars form in small haloes and/or multiple supernovae explode in each halo. 

For the interpretation of observed 21-cm absorption signals it will be of vital importance to understand the connection between the properties of Pop~III stars, their birth haloes and the recovery times. Once there is a reliable quantification of this connection, measurements of the Cosmic Dawn 21-cm signal with either radiometers or interferometers can be used to gain new constraints on the formation of the first stars and their environments. 

\section*{Acknowledgements}
The authors thank Tomoaki Ishiyama for providing the data from the Uchuu $N$-body simulations. AF was supported by the Royal Society University Research Fellowship. MM was supported by the Max-Planck-Gesellschaft via the fellowship of the International Max Planck Research School for Astronomy and Cosmic Physics at the University of Heidelberg (IMPRS-HD). 
Support for this work was provided by NASA through the Hubble Fellowship grant HST-HF2-51418.001-A, awarded by  STScI, which is operated by AURA, under contract NAS5-26555. 
SCOG, RSK and LHC, acknowledge funding from the Deutsche Forschungsgemeinschaft (DFG) via SFB 881 `The Milky Way System' (subprojects A1, B1, B2 and B8). TH acknowledge funding from JSPS KAKENHI Grant Numbers 19K23437 and 20K14464. This study is supported by the DFG via the Heidelberg Cluster of Excellence {\em STRUCTURES} in the framework of Germany’s Excellence Strategy (grant EXC-2181/1 - 390900948). The authors also gratefully acknowledge the data storage service SDS\@hd and the computing service bwHPC, supported by the Ministry of Science, Research and the Arts Baden-Württemberg (MWK) and the German Research Foundation (DFG) through grant INST 35/1314-1 FUGG. This research was supported for IR and RB by the Israel Science Foundation (grant No. 2359/20) and by the ISF-NSFC joint research program (grant No. 2580/17)

\section*{Data Availability}
The data underlying this article will be shared on reasonable request to the corresponding author.

\bibliographystyle{mnras}
\bibliography{all}

\appendix
\section{\textsc{a-sloth} model with feedback}
\label{apx:feed}
Both chemical and radiative feedback 
 crucially depend on the rate of Pop~III and Pop~II star formation in each modelled halo, and therefore come with many more free parameters. We do not study the dependence of the resulting 21~cm signal on all of these parameters, but merely aim to test the difference between having and not having small-scale feedback. Therefore, we only test one set of parameters, which we adopt from \citet{Tarumi20}, who calibrated the high-redshift star formation module of \textsc{a-sloth} using the metallicity distribution function of the Milky Way and reproduce the stellar mass to halo mass ratio from \citet{GarrisonKimmel17}.

Pop~II star formation is implemented as a four-phase bathtub model with the baryonic matter in haloes cycling between hot and cold ISM, stars and outflows \citep[see][for details]{Tarumi20}. The haloes can enrich nearby haloes with metals via outflows \citep{Magg18}. Additionally, we model ionizing bubbles around the haloes with the implicit R-type ionization front scheme \citep{Magg18}. We describe the numerical method used for deciding whether a halo is inside one of these ionized or enriched regions in Appendix \ref{apx:num}.
 
We follow the same fitting procedure as in Section \ref{sect:fitting} to find the functional form of $f_{\rm II}$, the fraction of Pop~II star forming haloes. The best-fitting parameters of this model are shown in Table \ref{tab:best_feed}. A comparison to Table \ref{tab:best_fit} reveals that the differences caused by the additional feedback between the haloes is much smaller than the difference arising from the recovery time. In other words, at the high redshifts we are considering, how exactly ionizing radiation and SNe affect the immediate vicinity of the Pop~III stars in the local halo has a bigger impact on the transition to metal-enriched star formation than their effect on the IGM. Therefore, and for the sake of simplicity, we use the semi-analytical model without external enrichment and ionization feedback to predict the 21-cm signature.
 
 \begin{table}
  \begin{tabular}{llll}
Model&Fast&Inter.&Slow\\
$t_\mathrm{recov}$ (Myr) & 10 &30&100\\
$F_0$&0.382&0.447& 0.472\\
$A$&0.493&0.474& 0.487\\
$\Delta z$&4.96&4.64& 2.96\\
$a_1$&4.16&2.97& 2.03\\
$a_2$&26.6&21.2& 15.3\\
$a_3$&-8.41&-3.26& -1\\
$a_4$&-14.5&-12.7& -10.4\\
\end{tabular}
\caption{\label{tab:best_feed}As Table \ref{tab:best_fit} but for modelling in full physics mode.}
\end{table}
 
\section{Numerical method for assigning feedback}
\label{apx:num}
Determining whether a halo is inside any of the ionized or enriched regions in the most simple implementation requires us to compute distances from the halo to the centres of all bubbles and to compare the distance to the size of the bubble. As \citet{Visbal20} pointed out, this leads to the cost of the computations scaling as $\propto N_\mathrm{source}N_\mathrm{halo}\propto V^2$, where $V$ is the volume of the simulated region. In our case, for the lowest critical temperature and at redshift $z=11$, we have 70 million haloes in the box, 4 million of which are star-forming. This would mean we would have to compute up to 280 trillion pairwise distances per time-step, which, on the computer we use would take a very long time\footnote{We run these semi-analytical simulations on the `fat' nodes of the BwForCluster MLS\&WISO in Heidelberg. Each node has four Intel Xeon E5-4620v3 processors with a theoretical peak node performance of 320 GFLOP per second. If this performance was reached, computing and comparing the distances would take only one day. However, the task is, unless it is heavily optimized, memory-access limited. Each position and enriched and ionized radius would need to be loaded once per halo and time-step, taking at least one week if the theoretical maximum memory bandwidth of 272 GB per second were reached.} and make the parameter exploration we perform unfeasible. \citet{Visbal20} solve this problem by looking at chemical and radiative feedback on a three-dimensional grid which is constructed via fast-fourier-transforms of radiation- and enrichment-fields. We address the issue with a tree-based approach:
 
Our aim is to reduce the number of distances we need to compute. In order to do this, we sort all actively star-forming haloes into an oct-tree structure. Each star-forming halo is assigned to the smallest node that fully encompasses the ionized and the enriched region around the halo. The tree is chosen such that it is larger than the simulated box and it has a maximum depth of 20 levels. In contrast to a classical oct-tree, such as the ones often used for calculating gravitational interactions in large $N$-body simulations, haloes are not associated with the parent node of the one they have been assigned to. When checking whether a halo is, e.g., enriched, it is sufficient to see whether it is enriched by any halo associated with every tree-node the target halo is inside of. This leads to a result that is exactly identical to testing every pairwise combination but uses substantially less computation time. For the whole simulation with $\Tc = 1500$~K, the number of distances we need to compute is reduced by 99.95 per cent from $4\times 10^{15}$ to $2\times 10^{12}$, and it runs in three hours on 40 cores. While this type of external feedback is not used in the main study and only enters in Appendix \ref{apx:feed}, it was developed for this work and will be used in a variety of future applications.
 
\section{Fit residuals}
\label{apx:res}
\begin{figure*}
\includegraphics[width=\linewidth]{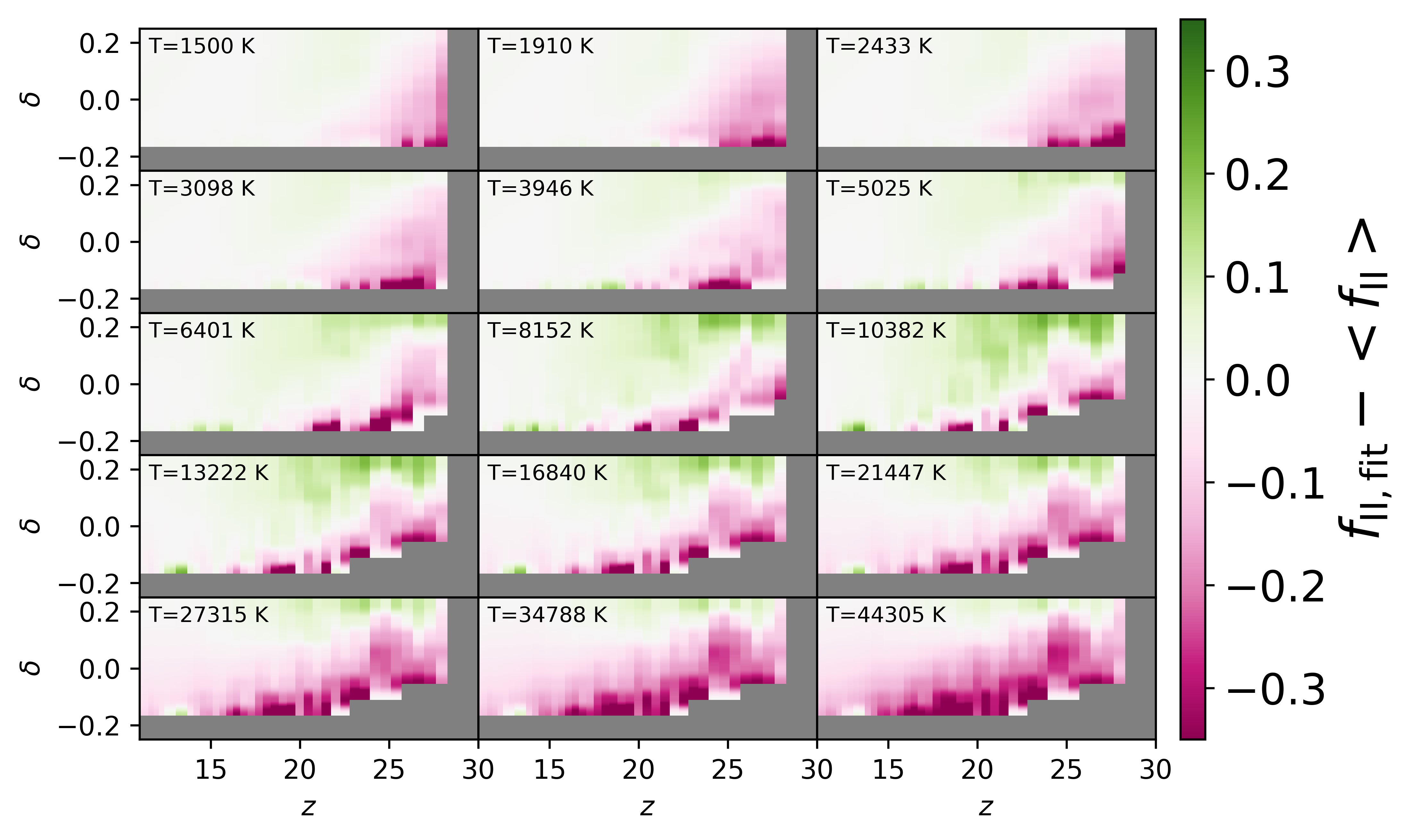}
    \caption{\label{fig:res10}Residuals for the fit of the Pop~II fraction for the slow transition. The area shaded in grey indicates no data or data that were excluded from the fits.}
\end{figure*}

\begin{figure*}
\includegraphics[width=\linewidth]{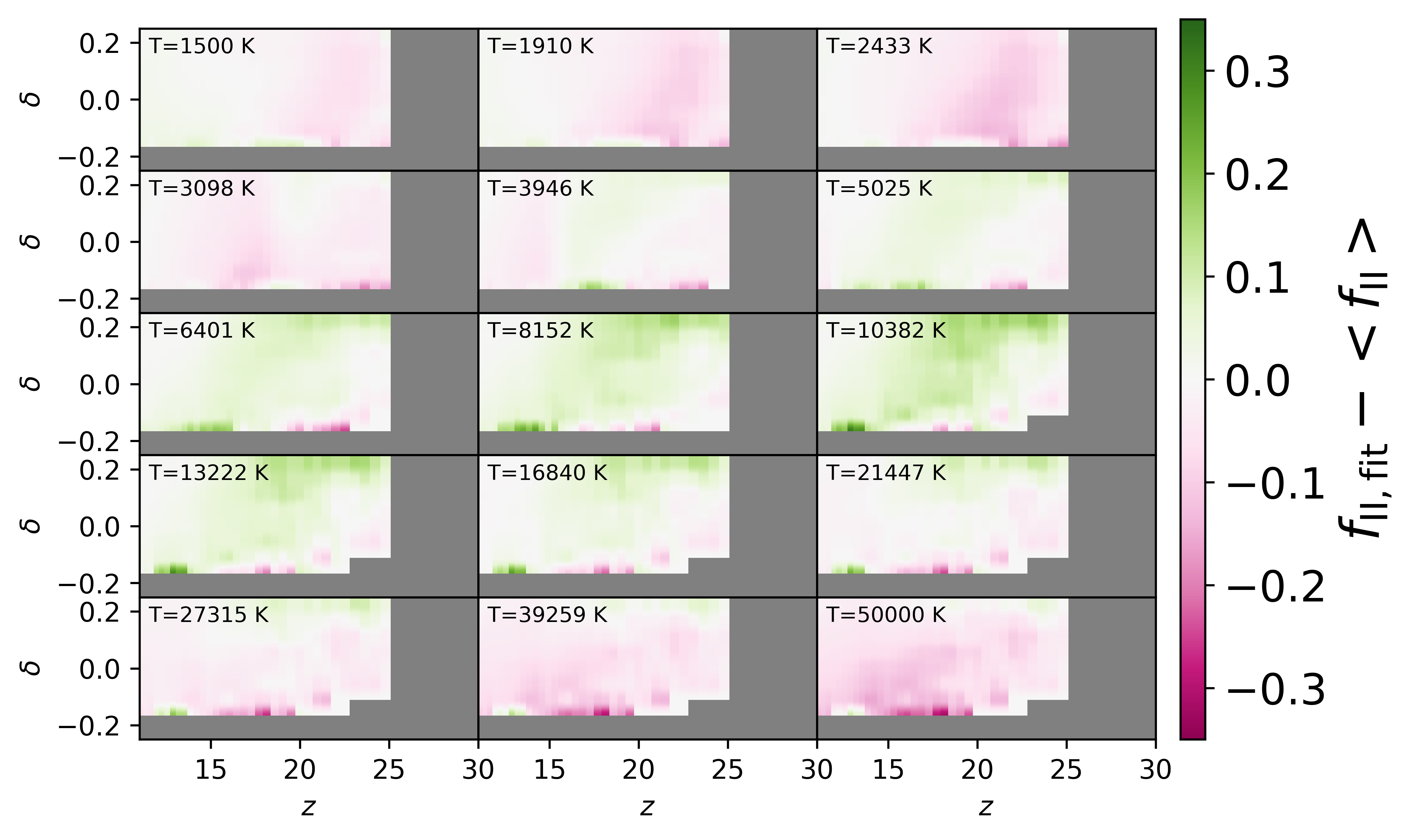}
    \caption{\label{fig:res30}Same as Fig. \ref{fig:res10} but for the intermediate transition.}
\end{figure*}

\begin{figure*}
\includegraphics[width=\linewidth]{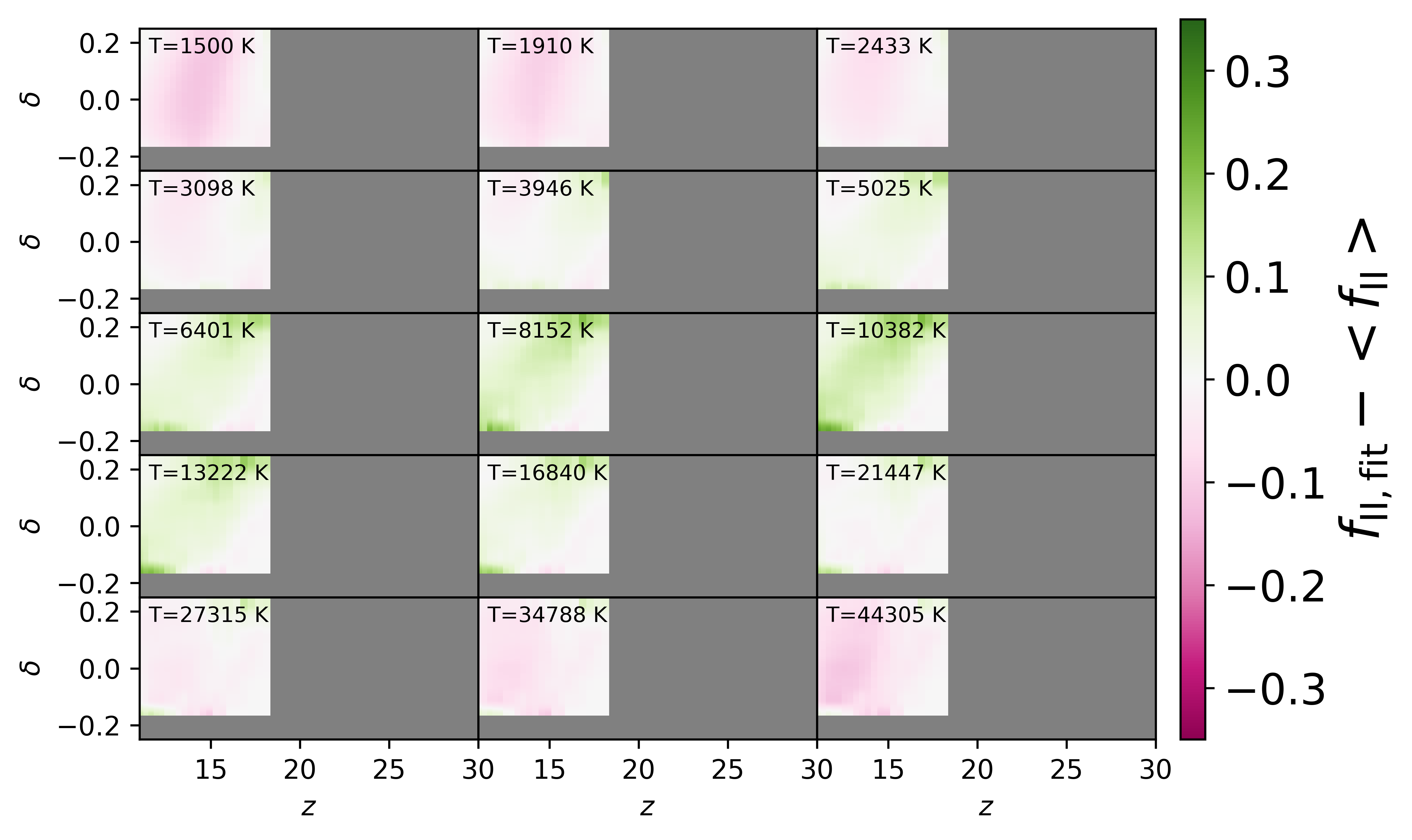}
    \caption{\label{fig:res100}Same as Fig. \ref{fig:res10} but for the slow transition.}
\end{figure*}

In this Section we show the quality of fit for the Pop~II fraction as discussed in Section \ref{sect:fitting}. For easier representation we only show the residuals for every second critical temperature. Figs. \ref{fig:res10}, \ref{fig:res30} and \ref{fig:res100} show the residuals for the fast, intermediate and the slow transition respectively. As we excluded the data obtained during the first fall-back time from the fits, the slow transitions have larger areas without data.

Generally the fits work well, with residuals of less than 10 per cent in most areas. The difference between the different transition speeds is much larger than uncertainties introduced by the fitting procedure. The residuals are largest in areas with low over-densities at high redshifts. These are the areas with the fewest star-forming haloes, which means that low-number statistics have a big impact here. However, as these are relatively rare low-density regions with below-average star-forming activity, we do not expect these regions to have a strong impact on the observed global 21cm signal.

\bsp 
\label{lastpage}
\end{document}